\documentclass[a4paper,fleqn,usenatbib]{mnras}
\usepackage[]{txfonts}
\usepackage{soul}
\usepackage{textcomp}
\usepackage{blindtext}
\usepackage{threeparttable}
\usepackage{graphicx,color}
\usepackage{epstopdf}
\usepackage{bm}
\newcommand{\be}{\begin{equation}}
\newcommand{\ee}{\end{equation}}
\newcommand{\eqb}{\begin{eqnarray}}
\newcommand{\eqe}{\end{eqnarray}}

\newcommand{\fermi}{{\sl Fermi}-LAT}
\newcommand{\sth}{\sigma_{\rm T}}
\newcommand{\spp}{\sigma_{\rm pp}}
\newcommand{\mel}{m_{\rm e}}
\newcommand{\mpr}{m_{\rm p}}
\newcommand{\rin}{r_{\rm i}}
\newcommand{\rout}{r_{\rm o}}
\newcommand{\rdec}{r_{\rm dec}}
\newcommand{\rw}{r_{\rm w}}
\newcommand{\vw}{v_{\rm w}}
\newcommand{\tw}{t_{\rm w}}
\newcommand{\tin}{t_{\rm i}}
\newcommand{\mcsm}{M_{\rm csm}}
\newcommand{\mej}{M_{\rm ej}}

\newcommand{\vsh}{v_{\rm sh}}
\newcommand{\no}{n_{\rm i}}
\newcommand{\eB}{\epsilon_{\rm B}}
\newcommand{\epr}{\epsilon_{\rm p}}
\newcommand{\dotn}{\dot{n}_{\rm IIn}}
\newcommand{\tpp}{t_{\rm pp}}
\newcommand{\kpp}{\kappa_{\rm pp}}
\newcommand{\gpmax}{\gamma_{\rm p,\max}}
\newcommand{\gpmin}{\gamma_{\rm p,\min}}

\title[Neutrinos from Type IIn supernovae]
{Point-source and diffuse high-energy neutrino emission from Type IIn supernovae}
\author[Petropoulou et al.]
{M. Petropoulou$^{1}$\thanks{Email: mpetropo@purdue.edu}, S. Coenders$^2$\thanks{Email: stefan.coenders@tum.de}, G. Vasilopoulos$^3$,  A.Kamble$^4$, L. Sironi$^5$\\ 
$^{1}$Department of Physics and Astronomy, Purdue University, 525 Northwestern
Avenue, West Lafayette, IN 47907, USA \\
$^2$ Technische Universit{\"a}t M{\"u}nchen, James-Frank-Str. 1, D-85748 Garching
bei M{\"u}nchen, Germany \\
$^3$ Max-Planck-Institut f\"ur extraterrestrische Physik,Giessenbachstra{\ss}e, 85748 Garching, Germany \\
$^4$ Harvard-Smithsonian Center for Astrophysics, 
60 Garden Street, Cambridge, MA 02138, USA \\
$^5$ Department of Astronomy, Columbia University, 550 W 120th St, New York, NY 10027, USA
}

\begin{document}
\date{Received / Accepted}
\pagerange{\pageref{firstpage}--\pageref{lastpage}} \pubyear{2016}

\maketitle

\label{firstpage}

\begin{abstract}
Type IIn supernovae (SNe), a rare subclass of core collapse SNe, explode in dense
circumstellar media that have been modified by the SNe progenitors at their
last evolutionary stages.  The interaction of the freely expanding SN ejecta
with the circumstellar medium gives rise to a shock wave propagating in the
dense SN environment, which may accelerate protons to multi-PeV energies.
Inelastic proton-proton collisions between the shock-accelerated protons and
those of the circumstellar medium lead to multi-messenger signatures. Here, we
evaluate the possible neutrino signal of type IIn SNe and compare with IceCube
observations.  We employ a Monte Carlo method for the calculation of the
diffuse neutrino emission from the SN~IIn class to account for the spread in
their properties. The cumulative neutrino emission is found to be $\sim 10$ per
cent of the observed IceCube neutrino flux above 60~TeV. Type IIn SNe would be the
dominant component of the diffuse astrophysical flux, only if 4 per cent of all
core collapse SNe were of this type and 20 to 30 per cent of the shock energy was
channeled to accelerated protons. Lower values of the
acceleration efficiency are accessible by the observation of a single type IIn SN as
a neutrino point source with IceCube using up-going muon neutrinos. Such an
identification is possible in the first year following the SN shock breakout
for sources within 20~Mpc.
\end{abstract}


\section{Introduction}
\label{sec:intro}
The identification of high-energy ($E_{\nu}>25$ TeV) neutrino sources would
provide direct evidence for the acceleration of cosmic rays (CR) up to $\sim$PeV
energies.  IceCube recently observed astrophysical neutrinos in both starting
event and up-going muon neutrino samples \citep{Aartsen13, Aartsen:2013jdh,
Aartsen:2016xlq}, but no significant anisotropies are identified in the
arrival directions of neutrinos. Additional searches for the origin of these
neutrinos have not yet revealed any specific sources
\citep{Adrian-Martinez:2014wzf, Aartsen:2016oji}. No prompt emission of
neutrinos was identified coincident with gamma-ray bursts
\citep{Aartsen:2016qcr} and no more than $27\%$ of the sub-PeV neutrino signal
may originate from blazars \citep{Aartsen:2016lir}, a type of radio-loud active
galactic nuclei with relativistic jets pointing towards the observer
\citep{blandfordrees78, urrypadovani95}. A dominant contribution of blazars beyond PeV energies can now also be excluded 
\citep[see e.g.][]{murase_waxman16}, although
a $10\%-20\%$ contribution is still viable \citep{padovani15, Aartsen16_10pev, padovani16}. 
Star-forming galaxies (i.e., galaxies with vigorous star
formation and high gas densities in their central regions) cannot contribute
more than $\sim30\%$ to the diffuse neutrino background between 25 TeV and 2.8 PeV  \citep{bechtol15}, 
if recent constraints from the non-blazar
extragalactic $\gamma$-ray background (EGB) \citep[][]{ackermann16} are taken into
account and the 30 TeV excess of IceCube events is attributed to star-forming galaxies \citep{aartsen15_combined}. 
Meanwhile, most of the scenarios predicting a
dominant Galactic contribution to the high-energy IceCube signal are disfavored
\citep[e.g.][]{ahlers16}.

There is convincing evidence that CR with energies up to the {\sl
knee} of the CR spectrum are accelerated at Galactic supernova (SN) remnants
\citep[for reviews, see][]{bell13,blasi13}. Acceleration beyond PeV energies
may be possible at shocks of interaction-powered SNe, i.e., SNe exploding in
dense circumstellar media (CSM) \citep{katz11, murase11, cardillo2015, ZP16}.
Due to the presence of (multi-)PeV protons in dense environments,  inelastic
proton-proton (p-p) collisions with the non-relativistic protons of the shocked
CSM  may lead to interesting multi-messenger signatures, such as GeV
$\gamma$-ray emission, high-energy ($>100$ TeV) neutrino production, and radio
emission \citep{murase14, pks16, ZP16}. In contrast to $\gamma$-rays, which may
be attenuated via photon-photon absorption soon after the shock breakout, when
their production rate is higher \citep[e.g.][]{kantzas16}, neutrinos escape the
source unimpeded. In principle, neutrino detection from an individual
interaction-powered SN would serve as the smoking gun for CR shock
acceleration to PeV energies.

Signs of interaction between the SN ejecta and the CSM are observable on the
early-time  light curves and spectra  (hours to days after the explosion) of
type IIn supernovae (SNe~IIn), a rare subclass of core collapse (CC) SNe
\citep{schlegel90, filippenko97}. The high CSM densities needed to explain
their observational properties can be accounted for, if the SN~IIn progenitor
has undergone strong mass loss before its explosion \citep[for a review,
see][]{smith14}. The inferred mass-loss rates are typically higher than
$10^{-3}$~M$_{\sun}$~yr$^{-1}$ \citep[e.g.][]{salamanca98, chugai04, kiewe12,
chandra15}, but they may be as low as $10^{-5}$~M$_{\sun}$~yr$^{-1}$
\citep{crowther07}. SNe~IIn progenitors exhibit wide diversity not only in
their estimated mass-loss rates but also in their wind velocities. These
typically lie in the range $\vw\approx10-10^3$ km s$^{-1}$ 
\citep[e.g.][]{salamanca98, chevalier00}. Giant progenitors have slower and
denser winds compared to those those from more compact progenitors which have
faster and more tenuous winds. 

SNe~IIn pose an interesting alternative to existing scenarios for neutrino
production, as it has been discussed first by \citet{murase11} and more recently by
\citet{ZP16} -- henceforth ZP16. \citet{murase11} demonstrated that multi-TeV
neutrinos are detectable by a generic IceCube-like detector for SNe at
$\lesssim 20-30$~Mpc, if the cosmic ray acceleration efficiency is 10 per cent.
ZP16 calculated the diffuse neutrino emission from the SN~IIn population, by
solving the hydrodynamical equations for the evolution of the SN shock and
taking into account particle acceleration and the CR feedback on the shock
structure.  The method presented in ZP16 is better suited for a single source,
as particle acceleration at both SN shocks (forward and reverse) is treated in
more detail,  but it is impractical when applied to many sources. The diffuse
neutrino emission was, therefore, calculated by adopting the same physical
parameters for all SNe (e.g., CSM density, shock velocity, and others). In
particular, the CR acceleration efficiency was fixed to be 50 per cent for all
SNe shocks, which might be unrealistically high
\citep{caprioli_spitkovsky14b,park2015}.

In this study, we calculate the neutrino signal from SNe~IIn and discuss the
possibility of constraining the CR accelerated energy fraction  by means of
diffuse and point-source neutrino observations with IceCube. We employ a Monte
Carlo method for the calculation of the diffuse neutrino emission from the
SN~IIn class, in an attempt to incorporate the wide spread of their properties
into their cumulative emission. The parameter values, which are assigned to the
simulated sources, are randomly drawn from distributions that are motivated by
observations. For each simulated source, we solve the evolutionary equations
for the proton and neutrino distributions, under the assumption that protons
are accelerated at the SN forward shock and produce neutrinos via p-p
collisions with the non-relativistic protons of the shocked CSM. For the
calculation of the diffuse neutrino emission, we adopt the redshift evolution
of CC SNe as presented in \citet{hopkins_beacom06}. 

Interestingly, a SN~IIn was recently
discovered in a search with the Palomar Transient Factory (PTF) \citep{law09} following an
IceCube neutrino doublet trigger alert. However, there was no evidence for a physical
connection, as the detection was most likely coincidental \citep{Aartsen:2015trq}. Motivated by this observation, we complement our analysis
by investigating the possibility of detecting an individual SN~IIn as a
neutrino point source with through-going muons detected with IceCube in the TeV
range over a period of seven years~\citep{Aartsen:2016oji}. By comparing the
arrival times and directions of IceCube neutrinos with known SNe~IIn,  we find one
starting neutrino event in close spatial and temporal correlation
 with a close-by SN~IIn in the Southern Sky. This finding motivates follow-up searches with
up-going neutrino data from the neutrino telescope ANTARES \citep{Adrian-Martinez:2014wzf}.

This paper is structured as follows. In Section \ref{sec:model}, we describe the
theoretical framework and our methods. In Section \ref{sec:results}, we compute
the diffuse neutrino emission from the SNe~IIn class and compare it with
IceCube neutrino observations. We calculate the neutrino signal expected from
individual sources and discuss the possibility of neutrino detection with
IceCube. In Section \ref{sec:discuss}, we discuss our results and model caveats.
We also discuss the possible association between SN2011fh and an IceCube event
of the starting sample. We finally conclude in Section~\ref{sec:conclusions}.
Here, we adopt a cosmology with $\Omega_{\rm K}=0$, $\Omega_{\rm M}=0.31$,
$\Omega_{\Lambda}=0.69$ and $H_0=69.6$ km s$^{-1}$ Mpc$^{-1}$. 

\section{Model and Methods}
\label{sec:model}
The CSM is modelled as an extended shell with mass density
$\propto r^{-2}$
\citep[e.g.][]{chevalier82} and outer radius $\rw = \vw \tw$, where $\vw$ is
the expanding velocity of the material that has been ejected from the
progenitor star over a period $\tw$ \citep[for a recent review,
see][]{smith14}.  The interaction of the freely expanding SN ejecta with the
CSM gives rise to a fast shock wave propagating in the CSM (forward shock) with
velocity $\vsh$ and a reverse shock that crosses the outer parts of the SN
ejecta.\footnote{We neglect the contribution of the reverse shock to the
neutrino emission, but we discuss our choice in Section \ref{sec:discuss}.} As
long as the interaction between the SN ejecta and the CSM takes place within a
region that is optically thick to Thomson scattering, $\tau_{\rm T} \gg 1$, the
SN shock is mediated by radiation, thus prohibiting particle acceleration
\citep[e.g.][]{murase11, katz11}. The radiation may escape when $\tau_{\rm
T}\sim c/\vsh$ \citep{weaver76}; this defines the so-called shock breakout time
$\tin$ and radius $\rin \sim \vsh \tin$ \citep[e.g.][]{ofek14}. For dense CSM
environments, as those considered in this work, the shock is expected to break
out in the wind, namely $\rin \gg  r_*$, where $r_*$ is the typical radius of
the stellar envelope.
 
Particle (electron and ion) acceleration can, in principle, take place at $r
\ge \rin$ after the SN shock becomes collisionless. Henceforth, we use  $\rin$
as the normalization radius. The CSM density profile may be written as 
\eqb 
n(r) = \no \left(\frac{\rin}{r}\right)^2 = \frac{K_{\rm w}}{m r^2},
\label{eq:density}
\eqe
where 
\eqb
\label{eq:Kw}
K_{\rm w} \equiv \frac{\dot{M}_{\rm w}}{4\pi \vw}
\eqe
is the wind mass loading parameter, $\dot{M}_{\rm w}$ is the  mass-loss rate of
the progenitor star, and  $m=1.4\, m_{\rm H}$ for a medium with 10 per cent He
abundance by number. 

The total CSM mass can be then estimated as 
\eqb
\mcsm =  4 \pi m \int_{r_*}^{\rw} \!\!\! {\rm d}r \, r^2 n(r) \simeq 4 \pi \no m \rin^2 \rw,
\label{eq:mcsm}
\eqe 
where $\rw \gg r_*$ was assumed.
Combining equation~(\ref{eq:mcsm}) with the condition $\tau_{\rm T}(\rin) =
c/\vsh$, where the Thomson optical depth is defined as $\tau_{\rm T}(r) =
\int_r^{\rw}{\rm d} y \, \sth n(y)$, we obtain $\rin$:
\eqb
\rin = \rw\left(1+\frac{4\pi m c\rw^2}{\sth \vsh \mcsm}\right)^{-1}.
\label{eq:rin}
\eqe
The CSM density at the shock breakout radius, $\no$, is then derived from
equations~(\ref{eq:mcsm}) and (\ref{eq:rin}) knowing $\rw$ and $\mcsm$. Its minimum
value, which is obtained when the shock breakout radius becomes maximum for
fixed CSM mass and shock velocity, is given by $\left(4 \pi^{1/3} m^{1/3} c/
\sth \vsh \mcsm^{1/3}\right)^{3/2}\simeq 9\times10^9\, {\rm cm}^{-3}
\left(\mcsm/10 M_{\sun}\right)^{-1/2} \left(\vsh/0.1 c\right)^{-3/2}$. 
We terminate our calculations at a maximum radius beyond which the contribution
to the total neutrino fluence is not important. This is set by the deceleration
radius
\eqb
\label{eq:rdec}
\rdec = \rin + \frac{\mej}{ 4\pi m \no \rin^2}
\eqe 
or by the extent of the CSM, i.e., $\rout=\min(\rdec, \rw)$. The production
rates of accelerated protons and neutrinos (at a fixed energy) are expected to
decrease beyond the deceleration radius, since they scale, respectively, as
$\vsh^3$ and $\vsh^2$ \citep[e.g.][]{pks16}. Furthermore, the neutrino
production rate is expected to decrease significantly beyond $\rw$, where the
density of the medium is much lower than $n(\rw)$. The condition $\rout=\rdec$
suggests that the ejecta mass is smaller than the total CSM mass, whereas
$\rout=\rw$ implies that $\mej > \mcsm$. 

To determine the temporal evolution of the neutrino emission from a single
source we calculate:
\begin{enumerate}
 \item the temporal evolution of the proton distribution from the shock
     breakout $\rin$ until the outer radius $\rout$.
 \item the maximum proton energy at each radius by comparing the local
     acceleration and loss timescales ($t_{\rm acc}$ and $t_{\rm loss}$). 
 \item the neutrino fluence $\phi_\nu$ (in erg cm$^{-2}$) by integrating the
     neutrino flux from the shock breakout till the time the shock decelerates
     or reaches the outer radius of the dense CSM.
\end{enumerate}
We solve the evolutionary equation for relativistic protons in the presence of
losses and injection. The latter is provided by Fermi acceleration
at the shock \citep[e.g.][]{mastichiadis96,kirkrieger98}. Neutrinos are
produced at a rate dictated by the relativistic proton distribution and the
non-relativistic proton density of the shocked CSM, which is assumed to be
uniform and equal to $4n(r)$ \citep[for details, see][]{pks16}, and they escape
from the shell without any losses. The equations for the proton and neutrino
distributions may be written as:
\eqb
\label{eq:proton}
\frac{\partial N_{\rm p}{(\gamma,r)}}{\partial r} + \frac{N_{\rm p}{(\gamma,r)}}{\vsh \tpp(r)} -\frac{\partial}{\partial \gamma}\left[\frac{\gamma}{r} N_{\rm p} {(\gamma,r)}\right]  
& = & Q_{\rm p}(\gamma, r)  \\  
\frac{{\rm d} N_{ {\nu_{\rm i}+\bar{\nu}_{\rm i}}}{(x_\nu,r)}}{{\rm d} r} + \frac{N_{{\nu_{\rm i}+\bar{\nu}_{\rm i}}}{(x_\nu,r)}}{\vsh t_{\rm esc}(r)}
& = & Q_{\nu_{\rm i}+\bar{\nu}_{\rm i}}(x_\nu, r)
\label{eq:neutrino}
\eqe
where $N_{\rm p}(\gamma,r)$ is the total number of protons in the shell with
radius $r$  with Lorentz factors between $\gamma$ and $\gamma+{\rm d}\gamma$,
$\tpp =\left(4 \kpp \spp c n(r) \right)^{-1}$ is the loss timescale for p-p
collisions with inelasticity $\kpp \approx 0.5$ and cross-section $\spp\simeq 3
\times 10^{-26}$~cm$^{-2}$,  $N_{\nu_{\rm i}+\bar{\nu}_{\rm i}}(x_\nu,r)$ is
the total number of neutrinos and anti-neutrinos\footnote{For simplicity, we
refer to the sum $\nu_{\rm i}+\bar{\nu}_{\rm i}$ as {\sl neutrinos}.} of
flavour i=e, $\mu$ with energies  (in $\mel c^2$ units) between $x_\nu$ and
$x_\nu+{\rm d}x_\nu$ at a shock radius $r$,  $t_{\rm esc} = h/c$, and $h\approx
r/4$ is the width of the shocked gas shell. Inelastic p-p collisions can be
treated as catastrophic energy losses \citep[e.g.][]{sturner97,schlickeiser02}
in contrast to the energy losses due to adiabatic expansion of the shell
(third term in the left hand side of equation~\ref{eq:proton}). Other energy
loss processes for protons, such as photopion cooling, are irrelevant \citep[see also][]{murase11}. 
The escape timescale of protons from the shell is assumed to be much longer than all other typical timescales of the system. Any decrease of the relativistic proton number due to the physical escape of the system would lead to lower neutrino fluxes than those presented in the following sections.

For a wind density profile, as adopted here, the proton injection rate is independent of radius
and is given by 
\eqb 
\label{eq:Qp}
Q_{\rm p}(\gamma)\equiv \frac{{\rm d}^2N_{\rm p}}{{\rm d}\gamma {\rm d}r}=\frac{9\pi}{8 f_{\rm p}}\epr \rin^2 \no \left(\frac{\vsh}{c}\right)^2 \gamma^{-p} H(\gamma-\gpmin)H(\gpmax-\gamma), 
\eqe
where $\epr$ is the fraction of the shock kinetic energy channeled to
accelerated protons, $f_{\rm p}=\ln\left(\gpmax/\gpmin\right)$ for power-law
index $p=2$ or $f_{\rm p}=(p-2)^{-1}$ for $p>2$, $\gpmin=1$, and $\gpmax$ is
the maximum Lorentz factor of the accelerated protons. This is determined by
$t_{\rm acc}=\min[\tpp, t_{\rm ad}]$, where $t_{\rm acc} \sim 6 \mpr \gamma
c^3/e B \vsh^{2}$ assuming Bohm acceleration. Here, $B=\sqrt{9 \pi \eB m \vsh^2
n}$, $\eB$ is the fraction of the post-shock magnetic energy, and $t_{\rm
ad}\sim r/\vsh$ corresponds to the source lifetime (see equations~(21)
and (29) in \citealt{pks16}). At small radii, where the CSM density is higher,
$\tpp < t_{\rm ad}$ leading to $\gpmax \propto r$. At larger radii, adiabatic
losses dominate and $\gpmax$ becomes independent of radius. The respective maximum neutrino energy is written as:
\eqb
\label{eq:Evmax}
E_{\nu, \max} = \left\{ \begin{array}{cc}
                       \frac{e B\vsh^2}{96 \spp n c^2} \propto \vsh^3 \eB^{1/2} K_{\rm w}^{-1/2} r & \tpp < t_{\rm ad} \\ \\
                       \frac{e B \kpp \vsh r}{24 c} \propto \vsh^2 \eB^{1/2} K_{\rm w}^{1/2} & \tpp  > t_{\rm ad}
                        \end{array}
\right.
\eqe 
The neutrino
production rate $Q_{\nu_{\rm i}+\bar{\nu}_{\rm i}}$ for parent proton energies
$E_{\rm p}>0.1$~TeV is written as \citep{kelner06}:
\eqb
\label{eq:qvm}
Q_{\nu_{\mu}+\bar{\nu}_{\mu}}(x_{\nu},r)\!\! \!\!& = & \!\! \!\!\frac{4 c n(r) \mel}{\vsh \mpr}\!\!\int_0^1 \!\!{\rm d}x \frac{\spp\left(E_{\rm p}\right)}{x}N_{\rm p}\left(\gamma, r\right)
\left(F_{\nu_\mu}^{(1)}+F_{\nu_\mu}^{(2)}\right), \\
Q_{\nu_{\rm e}+\bar{\nu}_{\rm e}}(x_{\nu},r)\!\! \!\!& = & \!\! \!\!\frac{4 c n(r) \mel}{\vsh \mpr}\!\!\int_0^1 \!\!{\rm d}x \frac{\spp\left(E_{\rm p}\right)}{x}N_{\rm p}\left(\gamma, r\right)
F_{\nu_{\rm e}},
\label{eq:qve}
\eqe
where $E_{\rm p}=\gamma \mpr c^2$, $x=x_{\nu}\mel c^2/E_{\rm p}$,
$F_{\nu_\mu}^{(1)}(x,E_{\rm p})$ and $F_{\nu_\mu}^{(2)}(x,E_{\rm p})\approx
F_{\nu_{\rm e}} $ are respectively given by equations~(66) and (62) in
\citet{kelner06}. For $E_{\rm p} \lesssim 0.1$ TeV, we adopt the
$\delta$-function approximation for the pion production rate as described in
\citet{kelner06}. Equations (\ref{eq:qvm}) and (\ref{eq:qve}) result in neutrino energy spectra that can be well described by 
a power law with index $\sim p$ for $E_{\nu} < E_{\nu, \max}$ and an exponential cutoff at $\sim E_{\nu, \max}$ (see also Fig.~12 in \citet{kelner06}).

The all-flavour neutrino energy flux from a source located at
a luminosity distance $D_{\rm L}$ is given by:
\eqb
E_\nu F_{\nu}(E_\nu, {r}) = \sum_{i=e,\mu} \frac{\mel c^2 x_{\nu}^2 N_{\nu_{\rm i}+\bar{\nu}_{\rm i}}(x_\nu,r)}{4\pi D_{\rm L}^2 t_{\rm esc}},
\label{eq:Fv}
\eqe 
where  $F_{\nu}$ is the differential neutrino energy flux (i.e., $F_{\nu}\equiv E_{\nu}{\rm d}N_{\nu}/{\rm d}E_{\nu}{\rm d}t$) at the shock radius $r$ and 
$E_{\nu}=x_{\nu} \mel c^2/(1+z)$ is the observed neutrino energy. The observed
neutrino fluence of a single source is 
\eqb 
\label{eq:fluence}
E_\nu\phi_{\nu}(E_\nu)=(1+z) \int_{\rin}^{\rout}{\rm d}r \frac{E_{\nu} F_{\nu}(E_\nu, r)}{\vsh}.
\eqe 
{
The diffuse all-flavour neutrino flux 
from the SNe~IIn class can be estimated as \citep[e.g.][]{cholis_hooper12}:
\eqb
\label{eq:diffuse}
E_\nu \Phi_{\nu}(E_\nu) \!\! & = & \!\!\frac{1}{4\pi}\int_0^{z_{\max}} \!\!\! {\rm d} z \frac{{\rm d}V_{\rm c}}{{\rm d}z} \, \frac{\dotn(z)}{1+z} \,E_\nu \phi_{\nu}(E_\nu),
\eqe
where ${\rm d}V_{\rm c}$ is the comoving volume element
\citep[e.g.][]{hogg99},  $z_{\max}=6$, $\dotn = \xi \,
\dot{n}_{\rm SNII}$ is the volumetric rate of SNe IIn, $\dot{n}_{\rm SNII}$ is the
volumetric rate of all CC SNe, and $\xi$ is the fraction of CC SNe that are of the IIn type. }
 \subsection{Physical parameters}
\label{sec:parameters}
The physical parameters of each simulated source which are randomly selected are summarized below:
\begin{enumerate}
 \item shock velocity,
 \item ejecta mass,
 \item duration $\tw$ of the pre-explosion period of mass loss,
 \item velocity $\vw$ of the wind, 
 \item total CSM mass, 
 \item fraction of the post-shock magnetic energy $\eB$, and 
 \item proton accelerated energy fraction $\epr$. 
\end{enumerate}
All other physical parameters of the system can be derived from a combination
of the above. For example, the extent of the CSM $\rw$ is determined by (iii)
and (iv). The shock breakout radius $\rin$ can be then estimated using
equation~(\ref{eq:rin}), if $\vsh$, $\rw$, and $\mcsm$ are defined. The last
two parameters incorporate the details of particle acceleration and magnetic
field amplification at the shock, while  parameters (i)-(v) are related to the
last stages of stellar evolution leading to the supernova explosion. 
For all the physical parameters mentioned above,
we adopt uniform distributions of random variables, as detailed in the following paragraphs. 
Given that the observations of SNe IIn are still not sufficient for pinpointing the distribution of their physical parameters, 
our choice ensures that our results will not be biased towards a more probable parameter set for the simulated sources.   

Kinetic simulations of particle acceleration at non-relativistic quasi-parallel
shocks (i.e., the angle between the shock normal and the magnetic field
direction is $\lesssim 45$\textdegree) have shown that $\epr=0.05-0.15$, while
$\epr \rightarrow 0$  for quasi-perpendicular shocks
\citep[e.g.][]{caprioli_spitkovsky14b}. As the proton accelerated energy
fraction depends on the  pre-existing magnetic field in the unshocked CSM, it
is appropriate to assume a range of values in our simulations. Adopting the
maximum $\epr$ predicted by theory would imply that the  unshocked CSM field in
SNe~IIn is weak or radial \citep[e.g.][]{sironi13}, which is questionable. In
our simulations we set $\log\epr = -1 -\tilde{r}$, where $\tilde{r}$ denote a
uniformly distributed random number in the range (0,1). 
 
The fraction of the post-shock magnetic energy is usually inferred from GHz
radio observations of interaction-powered SNe and is typically found to be
$10^{-3}-10^{-1}$ \citep{chevalier98, weiler02, chandra09, chandra15, kamble16}. 
Some of these estimates are obtained
assuming that the observed radio emission is produced by shock-accelerated
electrons and without taking into account electron cooling. The first
assumption may not be valid for the dense environments of SNe~IIn, where
relativistic electrons produced via the decay of charged pions from p-p
collisions, are expected to contribute to the observed radio emission
\citep{murase14, pks16}. Furthermore, electron cooling cannot be, in general,
neglected for the inferred $\eB$ values and the high CSM densities of
interaction-powered SNe \citep[see e.g.][]{marti-vidal11, kamble16, pks16}. It could be
therefore possible that $\eB \ll 10^{-3}$, as is usually the case for gamma-ray
burst afterglows \citep[e.g.][]{barniolduran14}. Because of the aforementioned
uncertainties, we choose a wide range of $\eB$ values, namely $\log\eB =
-5\tilde{r}-1$.

The shock velocity is one of the physical parameters that can be inferred from
optical spectroscopy and/or photometry. \citet{ofek14b} provided lower limits
for the shock velocity at the breakout by fitting the optical light curves of
15 SNe~IIn observed with PTF/iPTF \citep{law09}.  We created a sample of 23
sources by adding the shock velocity estimates of eight more sources that are
available in the literature (1986J, \citealt{bietenholz10}; 1994aj,
\citealt{salamanca98}; 1994W, \citealt{chugai04}; 1995G, \citealt{chugai03};
1995N, \citealt{chandra09}; 1997eg, \citealt{salamanca02}; 1997ab,
\citealt{salamanca98}; 2006jd, \citealt{chandra12}; 2010jl,
\citealt{chandra15}). The distribution of observed shock velocities (in
logarithmic space) is compatible with a uniform distribution with a median of
4300 km s$^{-1}$, although a Gaussian distribution cannot be excluded. Moreover, most of the values
in the observed sample are lower limits \citep[e.g.][]{ofek14b}. For these reasons, we have adopted a uniform distribution with a median of 9500 km s$^{-1}$
(i.e., $\log(\vsh/c)=-1-\tilde{r}$) in our simulations.

All other parameters are less well constrained observationally. The ejecta mass
typically ranges from 2-5 M$_{\sun}$ but it may also be as high as 15
M$_{\sun}$ depending on the progenitor type and stellar mass  at the zero age
main sequence (ZAMS) \citep[e.g.][]{shussman16}. Here, we adopt a uniform
distribution for the ejecta mass between 5~M$_{\sun}$ and 15~M$_{\sun}$, noting
that $\mej$ does not have a strong effect on the neutrino emission, as it
affects only the deceleration radius. There is even larger uncertainty on the
distribution of the CSM masses \citep[e.g.][]{smith14, margutti17}.
To account for the large spread in the estimated values, in
our simulations we assume a uniform distribution (in linear space) between 
0.01~M$_{\sun}$ and 20~M$_{\sun}$. 

The inferred wind velocities also display a diversity, i.e., $\vw\simeq
10-10^3$~km s$^{-1}$ \citep[e.g.][]{salamanca98, chevalier00}. Finally, $\tw$
is the least certain parameter, as it can be measured only for isolated cases,
namely $\eta$ Carinae, P Cygni and SN 2009ip \citep[e.g.][]{mauerhan13,
smith14}. The duration of the mass-loss events prior to explosion may typically
last from a few years up to several decades. Dynamical ages of shells of matter
around luminous blue variable stars (LBV) -- likely progenitors of SNe IIn
--  range from hundreds years to several thousand years \citep[see][and
references therein]{smith14}. These are, however, only indicators of the
mass-loss duration during the LBV phase. The uncertainty on $\tw$ is even
larger, as this may depend on the progenitor type and its initial mass. In our
simulations we, therefore, assume a wide range of values $\log\tw=1+2\tilde{r}$ (in
yr), while noting that the diffuse neutrino flux is not sensitive on the
maximum adopted duration. The $\tw$-distribution in combination with the
uniformly distributed $\vw$, results in $\rw$ values that are distributed
around $3\times10^{16}$~cm.

\begin{figure}
\centering
  \includegraphics[width=0.49\textwidth]{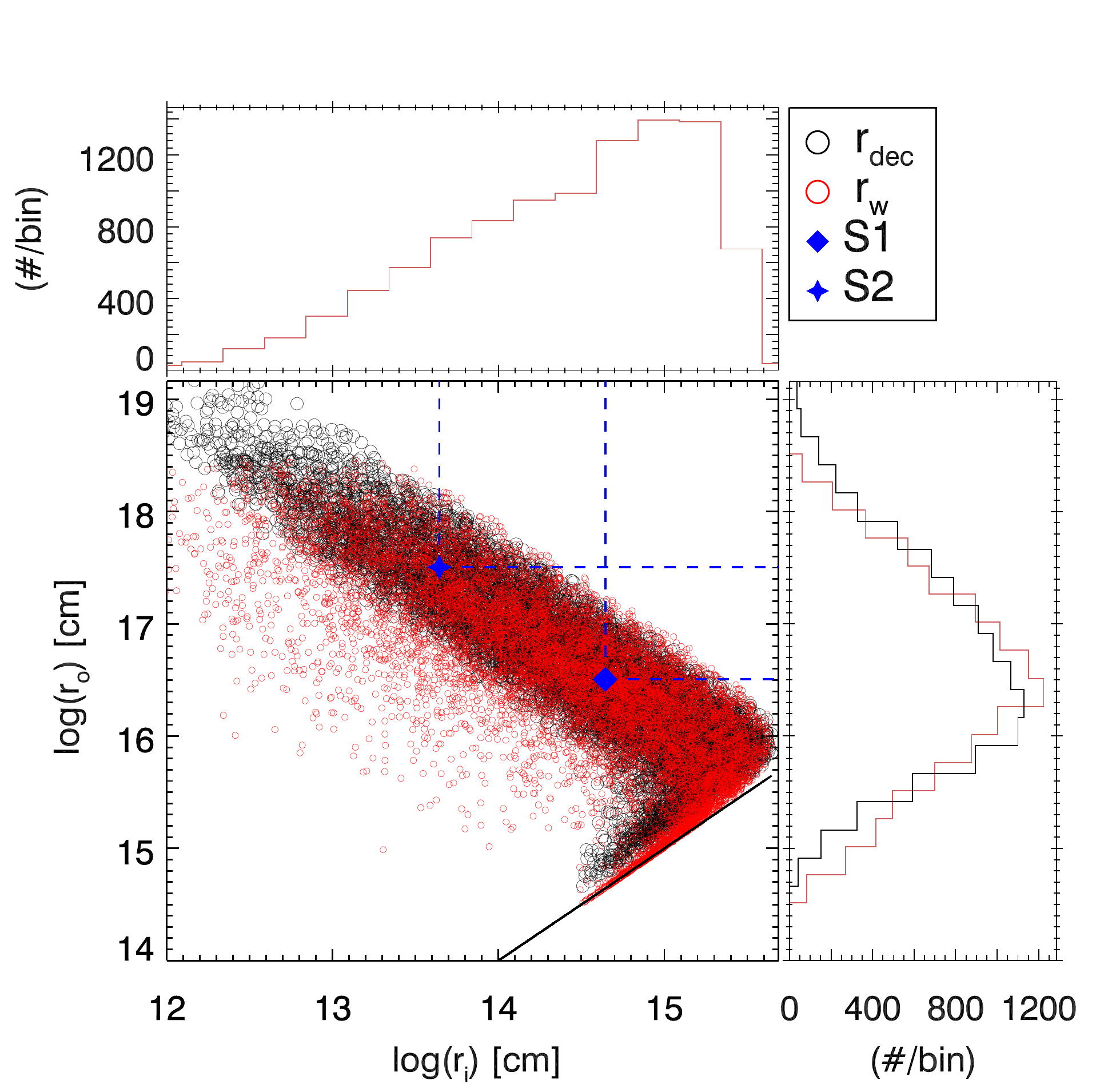}
 \caption{Log-log plot of the deceleration radius $\rdec$ (black points) and
 the CSM outer radius $r_{\rm w}$ (red points) versus the inner radius $\rin$ as
 derived from equation~(\ref{eq:rin}) for $10^4$ simulated sources. The
 parameter values $\left(\rin, r_{\rm w} \right)$ for the two scenarios
 discussed in Section \ref{sec:pointsource} are indicated with blue symbols.
 The density histograms of $\log\rin$ (top panel) and $\log\rw$, $\log\rdec$
 (right panel) are also shown.}
 \label{fig:rmax_rin}
\end{figure}
\begin{figure}
\centering
\includegraphics[width=0.49\textwidth]{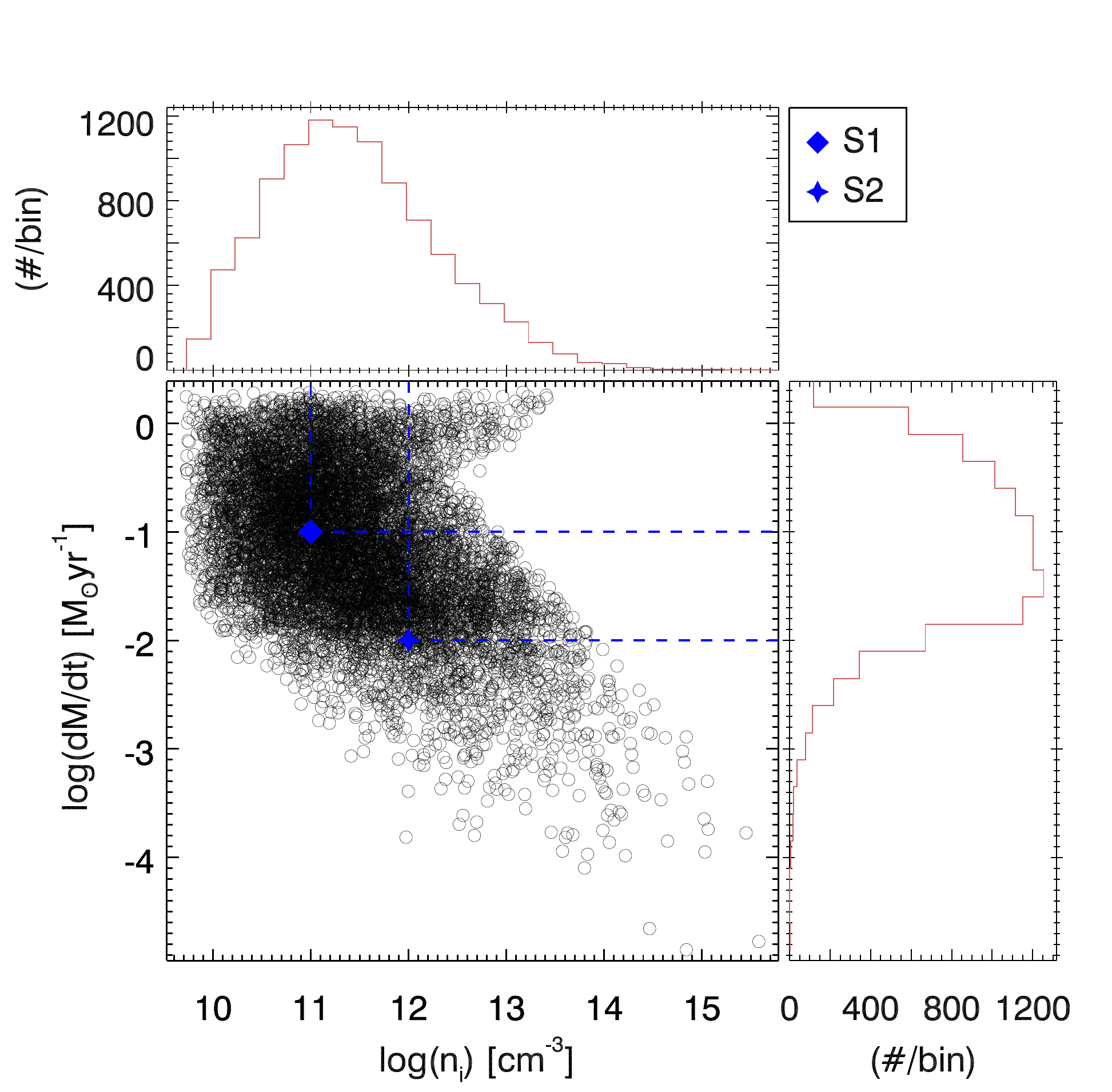}
 \caption{Log-log plot of the mass loss rate $\dot{M}_{\rm w}$ (black points)
 versus the CSM density at the shock breakout radius $\no$ for $10^4$ simulated
 sources. The parameter values for the two scenarios discussed in Section
 \ref{sec:pointsource} are indicated with blue symbols. The density histograms
 of $\log\no$ (top panel) and $\log\dot{M}_{\rm w}$ (right panel) are also
 shown.}
 \label{fig:dens_mdot}
\end{figure} 

\begin{figure}
\centering
\includegraphics[width=0.49\textwidth, trim=10 20 0 0]{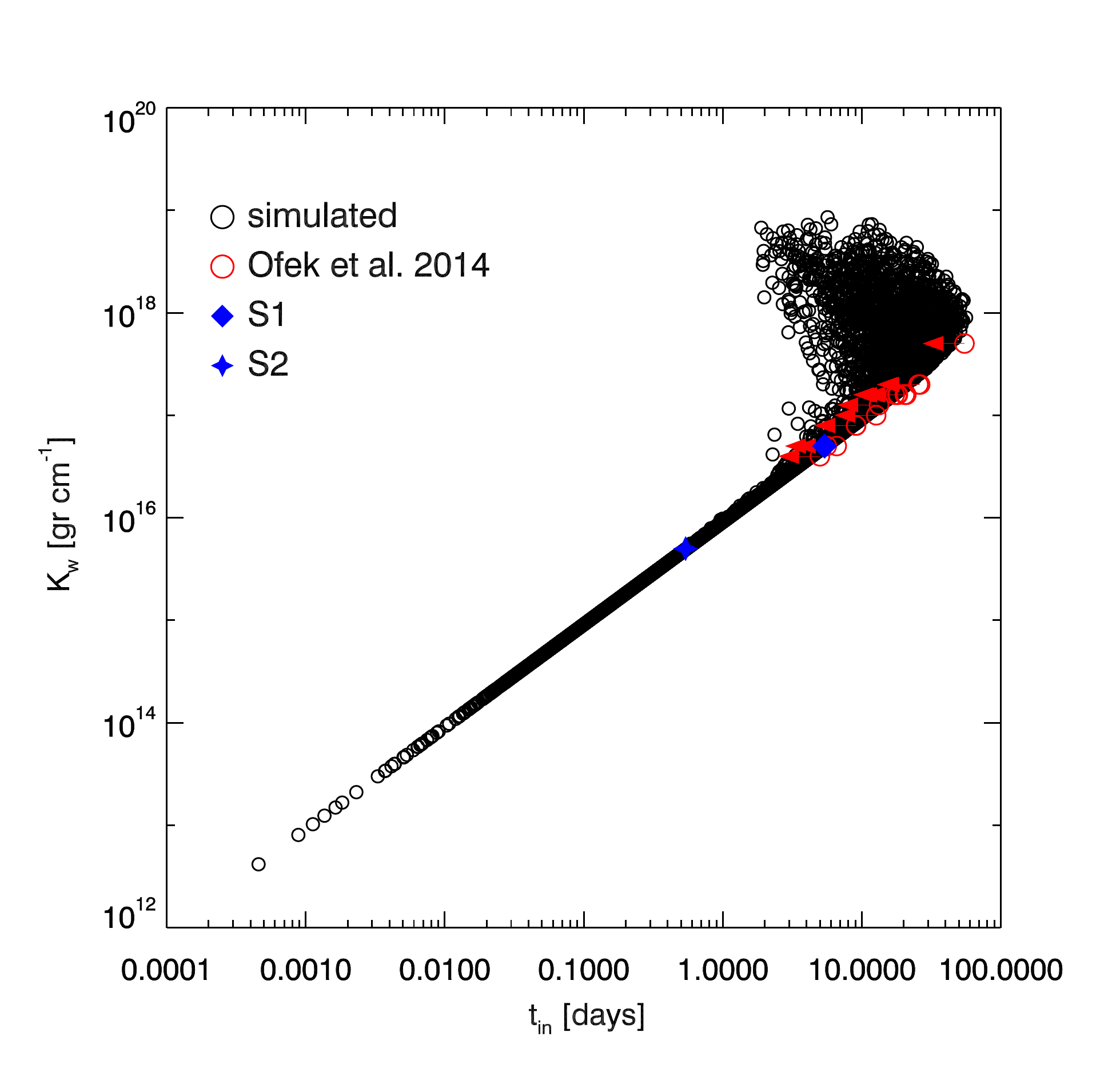}
 \caption{Plot of the mass-loading parameter $K_{\rm w}$ versus the shock
 breakout time used in our simulations (black points). The parameter values for
 the two scenarios discussed in Section \ref{sec:pointsource} are indicated
 with blue symbols. Values inferred by fitting the optical light curves of SNe IIn are overplotted with red symbols. The arrows indicate that
 these should be considered as upper limits of the actual shock breakout time.
 Data are adopted by \citet{ofek14b}. }
 \label{fig:Kw_tin}
\end{figure} 

Figure \ref{fig:rmax_rin} shows $\rdec$ (black points) and $\rw$ (red points)
versus $\rin$ for the simulated sources. No points lie below the solid line, as
expected, since this region corresponds to $\rin > \rout$. The boomerang-like
shape of the distribution of points in the middle panel can be understood after
inspection of equations~(\ref{eq:rin}) and (\ref{eq:rdec}). The parameter values for
the two scenarios discussed in Section \ref{sec:pointsource} are indicated with
blue symbols. The mass loss rate, defined as $\dot{M}_{\rm w}=4\pi \no m \rin^2
\vw$, is plotted against the CSM density at the shock breakout $\no$ in
Fig.~\ref{fig:dens_mdot}. Blue symbols have the same meaning as in
Fig.~\ref{fig:rmax_rin}. The median of density and mass loss rate distributions
are respectively $2\times10^{11}$~cm$^{-3}$ and $0.08$~M$_{\sun}$ yr$^{-1}$.
The wind mass-loading parameter, $K_{\rm w} =\dot{M}_{\rm w}/4\pi \vw$, is
plotted against $\tin = \rin/\vsh$ in Fig.~\ref{fig:Kw_tin} together with
estimates of the breakout time and the mass-loading parameter adopted by
\citet{ofek14b} (red symbols). The two scenarios discussed later in Section
\ref{sec:pointsource} are also shown as blue symbols. The golf club shape that
appears in the $K_{\rm w}-\tin$ plot results from the definition of $\rin$ and
$\no$. For small shock breakout radii, equation~(\ref{eq:rin}) reads $\rin \approx
\sth \vsh \mcsm/(4\pi m c \rw)$ and $\no \approx c/\sth \vsh \rin$. In this
limit, $K_{\rm w}\approx 4\pi m c \tin/\sth$ (golf club's shaft). At larger
shock breakout radii where $\rin \approx \rw$, $\no\approx \mcsm/4\pi m\rin^3$
and $K_{\rm w} \approx \mcsm/\vsh \tin$ (golf club's head) with the dispersion
arising from the random  $\mcsm$ and $\vsh$ values.

\subsection{Monte Carlo simulations}
We created $10^4$ parameter sets as described in the previous section and
calculated the respective neutrino emission. A total of $10^5$ random redshifts
($0 \le z\le z_{\max}$) were generated according to the distribution  ${\rm
d}V_{\rm c} \, \dotn(z) \, (1+z)^{-1}$. Henceforth, we adopt the CC SNe volumetric rate of \citet{hopkins_beacom06} and $\xi=0.04$ \citep[see e.g. Table~5 in][]{cappellaro15}.  Each of the
simulated sources was placed at 10 different redshifts and the diffuse neutrino
flux  
was  calculated as (see also equation (\ref{eq:diffuse})):
\eqb
E_\nu \Phi_{\nu}(E_\nu) = \frac{N_{\rm tot}}{4 \pi  N_{\rm sim}} \sum_{j=1}^{10}\sum_{i=1}^{10^4} E_\nu \phi_{\nu}^{(i,j)}(E_\nu),
\eqe
where $N_{\rm tot}=\left(10^{-7}\xi/3.1\right)\int_0^{z_{\max}} {\rm d}z
\left({\rm d}V_{\rm c}/{\rm d}z\right)\, \dotn(z) (1+z)^{-1}$ 
is the number of SNe~IIn per sec,   $N_{\rm sim}=10^5$ is the number of
simulated neutrino fluence spectra, and the indices i,j run over the different
parameter sets and redshifts, respectively.  Because of the dispersion
intrinsic to the simulation process, we repeated the above procedure for 100
different sets of redshifts. 

\section{Results}
\label{sec:results}
\subsection{Diffuse neutrino emission}
\label{sec:diffuse}
\begin{figure}
\centering
\includegraphics[width=0.49\textwidth, trim=15 10 0 0]{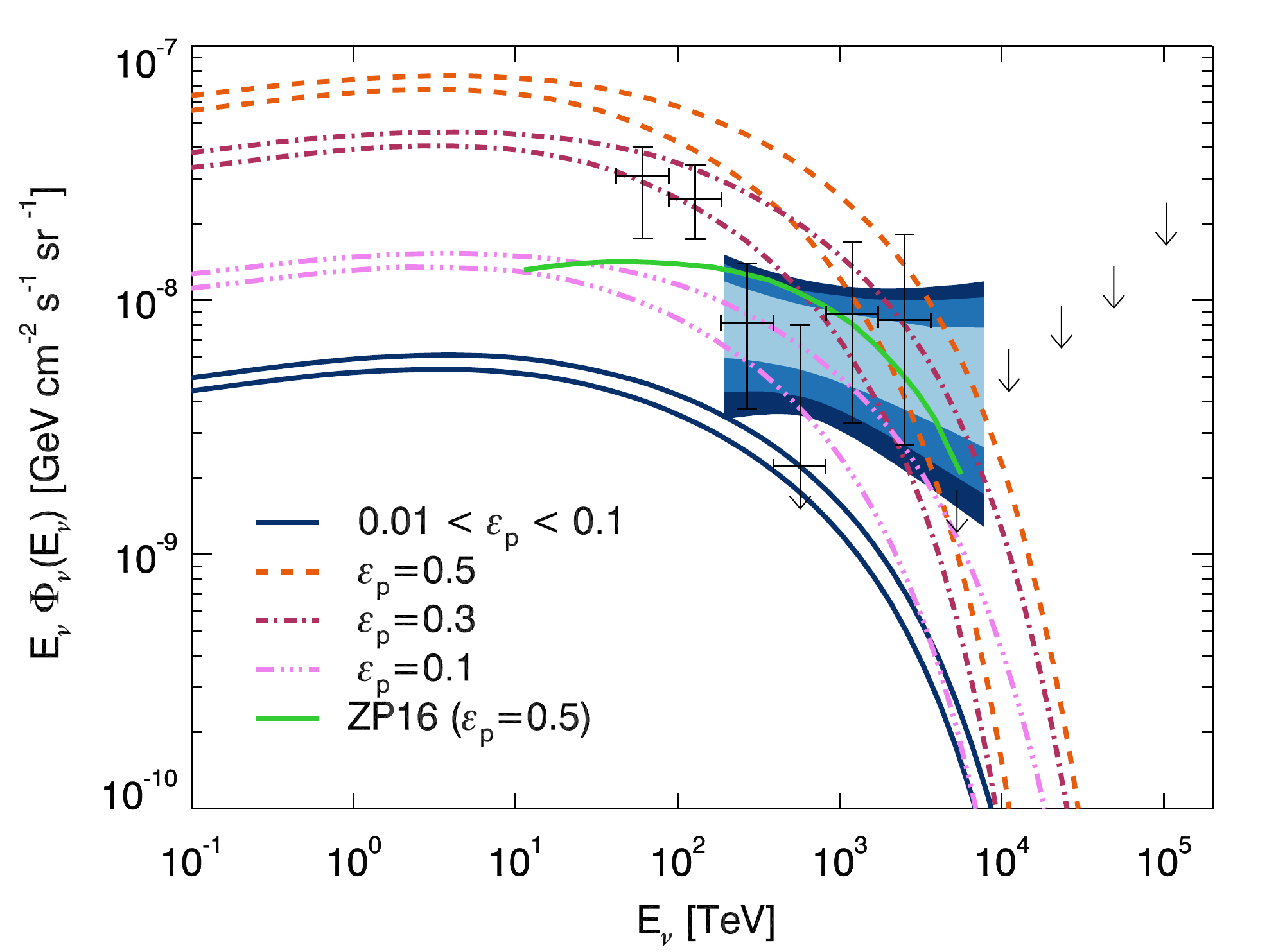}
\caption{Diffuse neutrino flux (per flavour) from SNe~IIn as obtained from our
    simulations when $\epr$ is a random number uniformly distributed between
    0.01 and 0.1 (solid blue line) or $\epr$ is fixed for all sources (see
    legend). For each case, the two curves enclose the 68 per cent of the
    simulated spectra ($\sim 1\sigma$ of a Gaussian distribution). It is
    assumed that SNe~IIn consist 4 per cent of the CC SNe population. The model
    prediction of ZP16 for $\epr=0.5$ and $\xi=0.01$ is overplotted for
    comparison. The points represent the differential energy spectrum of
    astrophysical neutrinos from the 4-year high-energy starting event (HESE)
    IceCube sample. The one standard deviation upper limits are shown as arrows
    \citep{ICRC15}. The best-fit neutrino spectrum of the 6-year
    IceCube muon neutrino sample from the Northern Hemisphere is overplotted (shaded band) for different uncertainties: 68 per cent (cyan), 95 per cent (light blue),
    and 99 per cent (dark blue) \citep{Aartsen:2016xlq}.} 
\label{fig:diffuse}
\end{figure}
Figure \ref{fig:diffuse} presents our predictions for the diffuse neutrino flux
(per flavour) from SNe~IIn in comparison with the astrophysical neutrino flux
detected by IceCube \citep{ICRC15, Aartsen:2016xlq}. The blue solid lines
enclose the 68 per cent of the simulated spectra when the accelerated proton
energy fraction varies among sources in the range $[0.01, 0.1]$. All other
curves are obtained under the assumption of a constant acceleration efficiency
among sources. 

The energy spectrum of the diffuse neutrino emission is 
best described by a power law with index $\sim 2$ and exponential cutoff at
$E_{\nu, \rm c} \sim 50$~TeV, which results from the convolution of individual neutrino spectra
with different cutoff energies (equation (\ref{eq:Evmax})). The dispersion of the rest frame cutoff energies is mainly driven by the wide range of $\eB$ values used in the simulations (see Section \ref{sec:parameters}). $E_{\nu, \rm c}$ would shift towards higher energies, if $\eB \ge 10^{-3}$  in all sources. The location of the exponential cutoff shown in Fig.~\ref{fig:diffuse} also depends on the assumption of Bohm acceleration. Were the acceleration process slower or suppressed \citep[e.g.,][]{murase11, metzger16}, the energy spectrum of the diffuse neutrino flux would steepen at a few TeV energy. 

The cumulative neutrino emission is $\sim 10$ per cent
of the observed IceCube neutrino flux above 60~TeV, if the accelerated proton energy fraction varies between 0.01 and 0.1 in different sources (solid blue lines). However, if $\epr=0.5$ for all SNe~IIn (i.e., higher
than most optimistic theoretical predictions), the model-predicted neutrino
flux (orange dashed lines) would exceed the observations. The IceCube flux could
be still explained by SNe~IIn with $\epr=0.5$, if the SNe~IIn fraction was one
per cent, as shown by ZP16 (green solid line).  Regardless, the 4-year HESE
IceCube sample already excludes the most extreme value for the proton
accelerated energy fraction, namely $\epr < 0.5$ for $\xi=0.04$. Similar
constraints on $\epr$ can be placed using the 6-year IceCube muon neutrino sample
from the Northern Hemisphere (shaded band in Fig.~\ref{fig:diffuse}). 
This sample has a higher energy threshold compared to the HESE sample
($\sim194$~TeV compared to $\sim60$~TeV, respectively), but higher precision in
the derived flux normalization and spectral index. At $E_{\nu}
\lesssim500$~TeV, the model predictions for $\epr=0.3$ and $\xi=0.04$ are in
conflict with the $99\%$ contours of IceCube. The tension with IceCube's 95\% confidence limits is resolved, if the
acceleration efficiency is as low as $(\epr/0.2) \times (\xi/0.04) \lesssim 1$.

 \begin{figure}
\centering
\includegraphics[width=0.49\textwidth,trim=15 10 0 0]{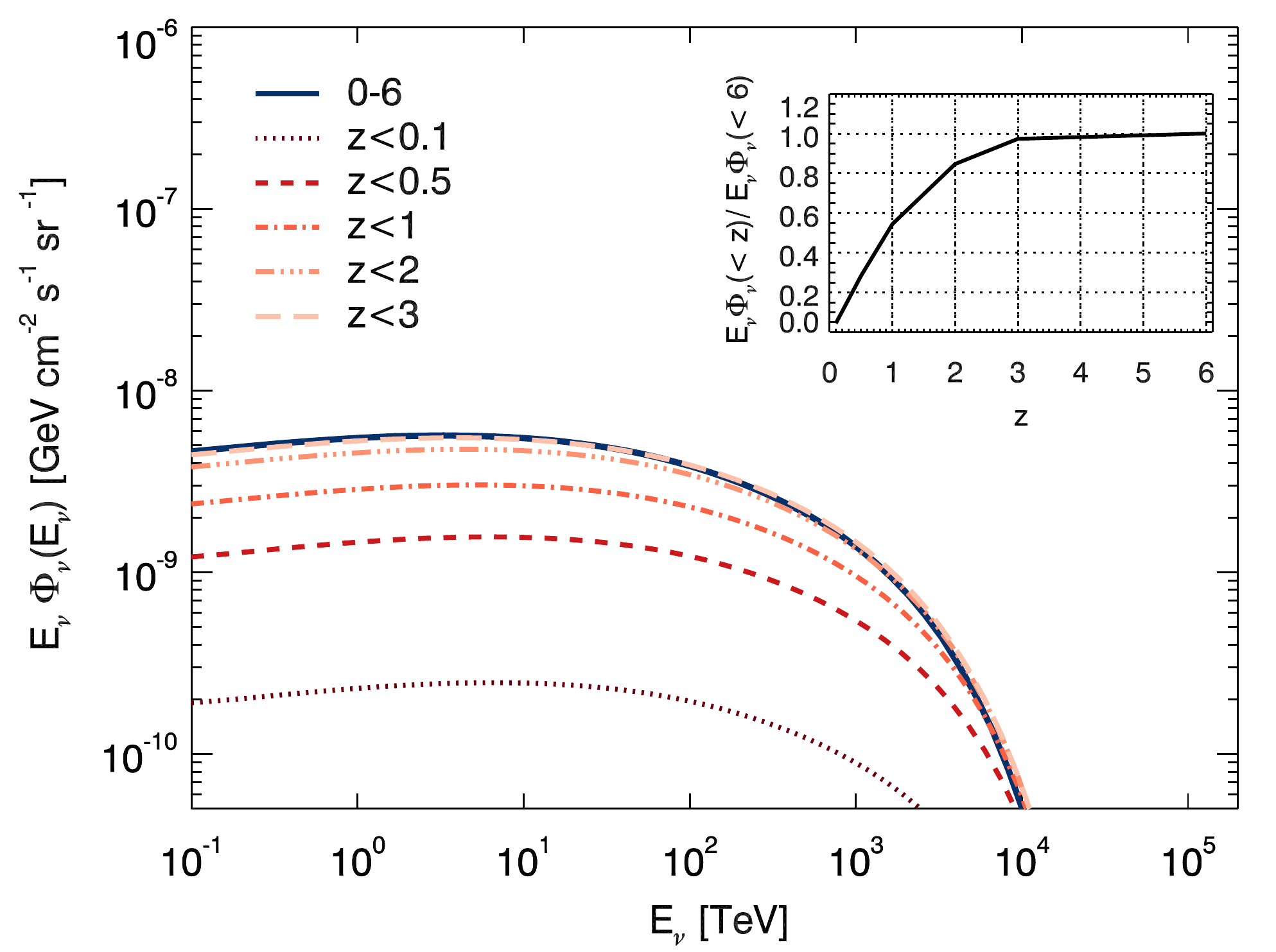}
\caption{Diffuse neutrino flux (per flavour) from SNe~IIn at different ranges
    of redshift. All curves correspond to the median spectrum of our
    simulations when $\epr$ is a random number uniformly distributed between
    0.01 and 0.1. The fractional contribution to the total diffuse neutrino
    flux from  SNe~IIn up to redshift $z$ is shown in the inset plot.  } 
\label{fig:diffuse_bins}
\end{figure} 

The contribution of SNe~IIn at different redshift ranges to the total diffuse
neutrino flux is illustrated in Fig.~\ref{fig:diffuse_bins}. Approximately 50
per cent of the total neutrino flux originates from sources at $z<1$ or $D_{\rm
L} <6.6$~Gpc. Sources at $1<z<2$ contribute about 30 per cent to the total
flux, while the contribution of all sources beyond $z=2$ drops to 20 per cent,
as a result of the decreasing volumetric production rate of CC SNe and
decreasing neutrino fluence from a single source, which cannot match the
increase of the comoving volume with redshift.
\begin{table}
    \centering
    \caption{Parameters used for the two SN scenarios S1 and S2.}
    \label{tab:case_parameters}
    \begin{tabular}{lcc}
        \hline
        Parameter            & S1                 & S2 \\
        \hline
        $\rw$ [cm]           & $3.2\times10^{16}$ & $3.2\times10^{17}$ \\
        $\mej$ [M$_{\sun}$]  & \multicolumn{2}{c}{10} \\
        $\mcsm$ [M$_{\sun}$] & \multicolumn{2}{c}{10} \\
        $\vsh$ [km s$^{-1}$] & \multicolumn{2}{c}{$9.5\times10^3$} \\
        $\eB$                & \multicolumn{2}{c}{$3\times10^{-4}$} \\
        $\epr$               & \multicolumn{2}{c}{$3\times10^{-2}$} \\
        $\rin$ [cm]          & $4.4\times10^{14}$ & $4.5\times10^{13}$ \\
        $\rout$ [cm]         & $3.2\times10^{16}$ &  $3.2\times10^{17}$\\
        $\no$ [cm$^{-3}$]    & $10^{11}$          & $10^{12}$ \\
        $K_{\rm w}$ [g cm$^{-1}$] & $4.5\times10^{16}$ & $4.5\times10^{15}$ \\
        $\dot{M}_{\rm w}$ [$M_{\sun}$ yr$^{-1}$]$^\dagger$ & 0.1 & 0.01 \\ 
        $\tin$ [d]            & $5.4$              & $0.54$ \\
        Duration [yr]        & $1$                & $10.7$ \\
        \hline
    \end{tabular}
    \\ $^\dagger$ A wind velocity $u_{\rm w}=100$ km s$^{-1}$ was assumed. \\
    Note --- All the parameters above $\rin$ are assigned to the median values
    of the respective distributions, except for $\rw$ in S2, which is ten times larger than the median value.
    All other parameters are  derived from the  preceding ones using equations~(\ref{eq:density})-(\ref{eq:rin}).
   \end{table} 
\subsection{SNe~IIn as neutrino point sources}
\label{sec:pointsource}
We present the evolution of the neutrino spectrum from a fiducial SN~IIn
located at $D_{\rm L}=10$~Mpc. We assume that neutrino production begins after
the shock breakout and terminates when the shock decelerates or reaches the
outer edge of the CSM extended shell. Neutrino production is expected to be
less important after this point because either the shock decelerates or it
encounters a sharp negative density gradient at $r\gtrsim \rw$.
\begin{figure*}
\centering
 \includegraphics[width=0.49\textwidth]{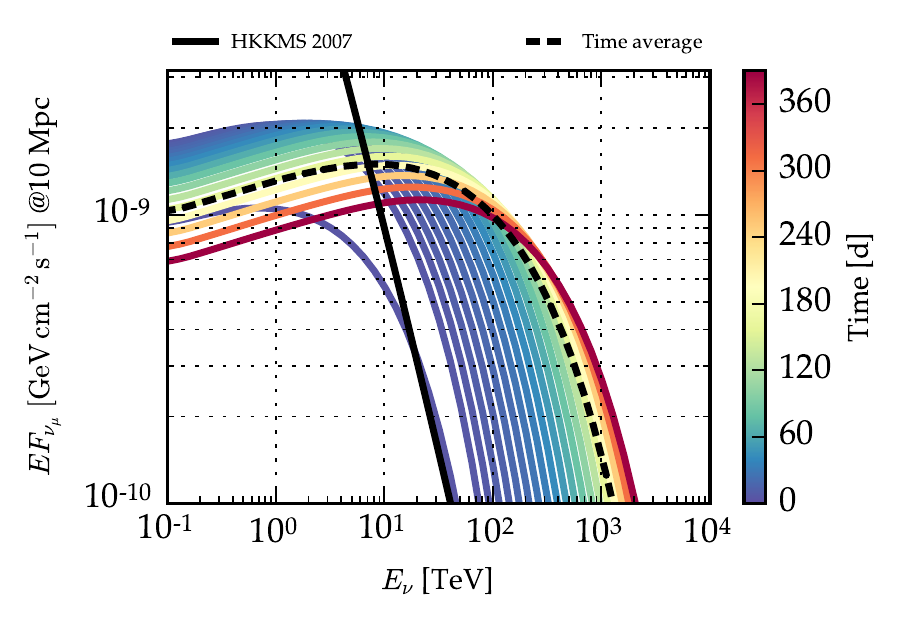}
 \includegraphics[width=0.49\textwidth]{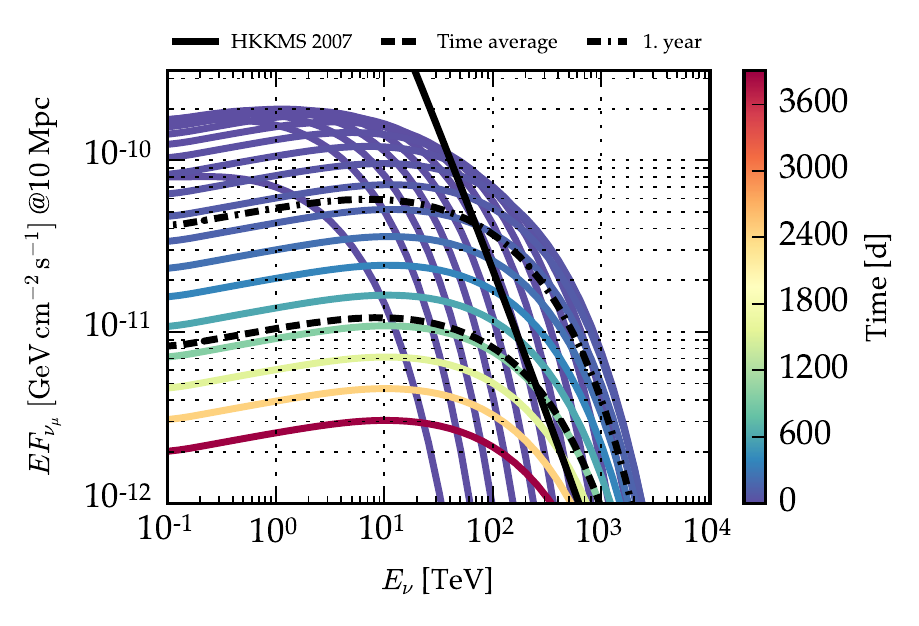}
  \caption{Snapshots of the $\nu_{\mu}+\bar{\nu}_\mu$ energy spectrum at
      different times (indicated by different colours) since the shock breakout
      until the shock radius reaches $\min[\rw, \rdec]$.  The fluxes are
      obtained after taking into account neutrino mixing due to oscillations.
      The energy spectrum of the atmospheric neutrino flux
  \citep{Aartsen:2015xup, Honda:2006qj} is overplotted (thick black line).
  Left- and right-hand panels show the results for S1 and S2, respectively.}
\label{fig:flux}
\end{figure*}

\begin{figure*}
\centering
 \includegraphics[width=0.49\textwidth]{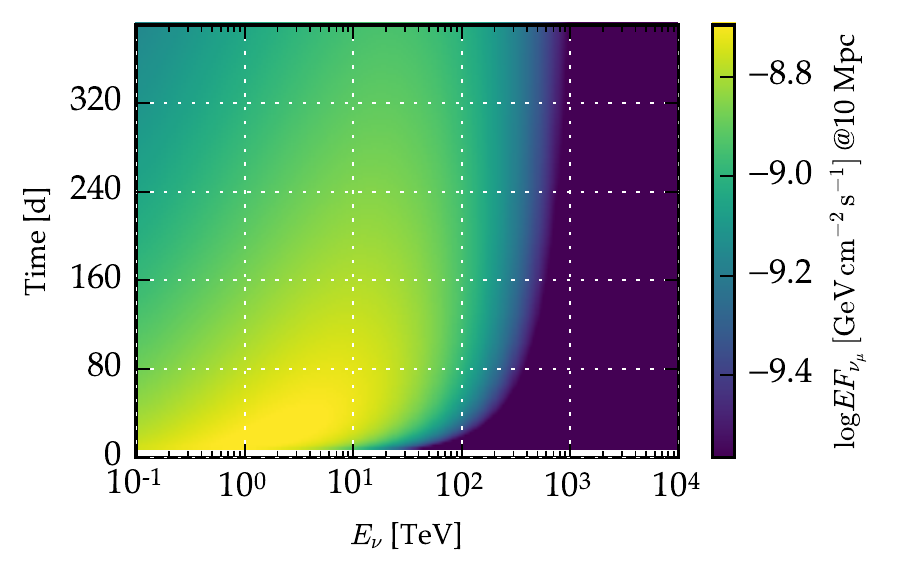}
 \includegraphics[width=0.49\textwidth]{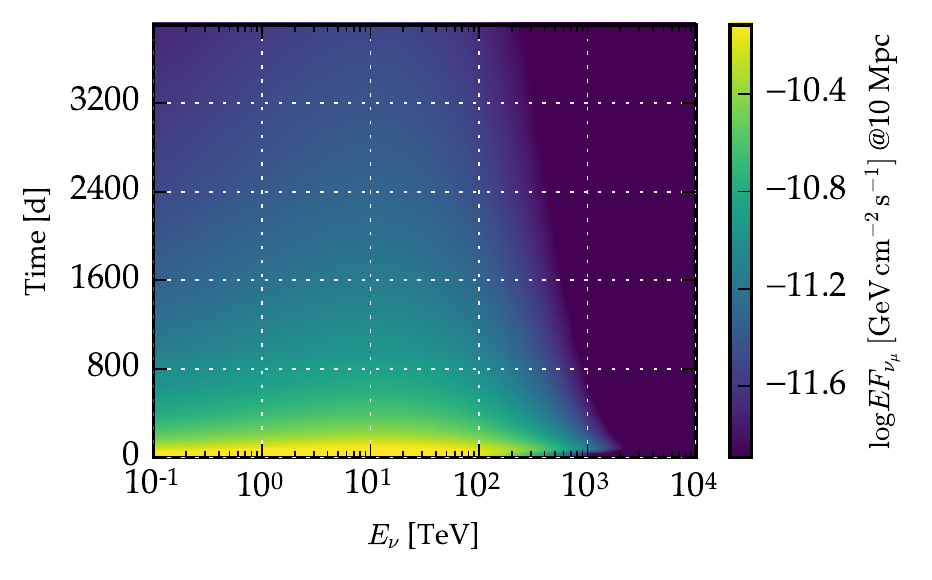}
 \caption{Two-dimensional plot of the $\nu_{\mu}+\bar{\nu}_\mu$ flux as a
 function of time at different energies for S1 (left-hand panel) and S2
 (right-hand panel).}
\label{fig:map}
\end{figure*} 

The neutrino flux is expected to reach its maximum value soon after the shock
breakout, with the exact peak time depending on the neutrino energy. This will
be exemplified below through numerical calculations. After the peak time, the
flux typically decays as a power law with time, i.e., $F_{\nu}(E_{\nu}) \propto
t^{-2+s}$. Although the exact value of $s$ depends on the model parameters, $s$
is bound between 1 and 2. The two limiting values are obtained in the following
regimes:
\begin{itemize}
 \item $s=1$, when proton injection is balanced by adiabatic losses. The
     neutrino flux then scales as:
 \eqb 
 \label{eq:F_adiab}
 F_{\nu}(E_{\nu})\propto \epr \vsh K_{\rm w}^2  t^{-1}.
 \eqe 
 \item $s=2$, when proton injection is balanced by p-p losses. The neutrino
     flux is given by \citep[see also][]{ZP16}:
 \eqb 
 \label{eq:F_pp}
 F_{\nu}(E_{\nu})\propto \epr \vsh^3  K_{\rm w} t^{0}.
 \eqe  
\end{itemize}
By comparing $t_{\rm pp}$ with the shock's dynamical timescale ($r/\vsh$), we
find that the second regime is relevant for
\eqb
\label{eq:time}
t\lesssim 100 \, {\rm d} \, K_{\rm w, 16}\,\beta_{\rm sh, -1.5}^{-2},
\eqe
where $\beta_{\rm sh}\equiv\vsh/c$.
 
We consider two scenarios (henceforth, S1 and S2) for the neutrino production.
Their parameter values used are listed in Table~\ref{tab:case_parameters}. In
S1 the parameters $\rw, \mej, \mcsm, \vsh,\eB,$ and $\epr$ were assigned to the median
values of the corresponding distributions (see Section~\ref{sec:parameters}). Thus, S1 is a representative scenario of the simulated SN~IIn population.
The impact of different parameters is illustrated via S2, where a ten times
larger $\rw$ was adopted, while $\mej, \mcsm, \vsh,\eB,$ and $\epr$ were kept
fixed (see Table~\ref{tab:case_parameters}). Hence, the shock breakout time is
$\tin=0.54$~d and the neutrino production lasts $\sim10.7$~yr.

Snapshots of the $\nu_\mu+\bar{\nu}_\mu$ energy spectrum at different times
following the shock breakout  are presented in Fig.~\ref{fig:flux} for S1
(left-hand panel) and S2 (right-hand panel). The results are obtained after
taking into account neutrino mixing due to oscillations. The soft energy
spectrum of atmospheric muon neutrinos \citep[HKKMS 2007 --][]{Honda:2006qj,
Aartsen:2015xup} is overplotted for comparison (thick black line).  The
neutrino flux increases rapidly at early times, but later decreases with a
slower rate, which depends on the specifics of the source. Meanwhile, the
maximum neutrino energy is increasing due to the increasing maximum energy of
the parent protons. This is evident in both cases during the first year. At
late times, where the adiabatic energy losses are more important than those
caused by p-p collisions, the maximum energy of protons  and, in turn,
neutrinos, remains constant.  Even in the optimistic scenario, where particle
acceleration proceeds at the fastest possible rate, the neutrino spectrum from
S1 barely extends beyond 1~PeV, as  shown in Fig.~\ref{fig:flux}. However,
stronger magnetic fields, faster shocks, and higher mass-loading parameters may
result in multi-PeV neutrino cutoff energies {(see equation (\ref{eq:Evmax}))}.

The temporal evolution of the $\nu_\mu+\bar{\nu}_\mu$ neutrino flux at
different energies is illustrated in Fig.~\ref{fig:map}. For S1, the peak flux
in the energy range 1-10 TeV is expected within the first 50 days after the shock
breakout. On the contrary, the 10-100 TeV flux remains approximately constant
for a long period lasting hundreds of days. No significant energy dependance of
the temporal evolution of the neutrino flux is found for S2, indicating an
approximately constant spectral shape (see also Fig.~\ref{fig:flux}). This, in
turn, implies that adiabatic expansion governs the temporal evolution of the
parent proton energy distribution. Substitution of the relevant $K_{\rm w}$ and
$\vsh$ values in equation (\ref{eq:time}) also shows that p-p collisions are
more important than adiabatic expansion only  for $t\lesssim 45$~d. 

\begin{figure}
\centering
\includegraphics[width=0.49\textwidth,trim=15 10 15 0]{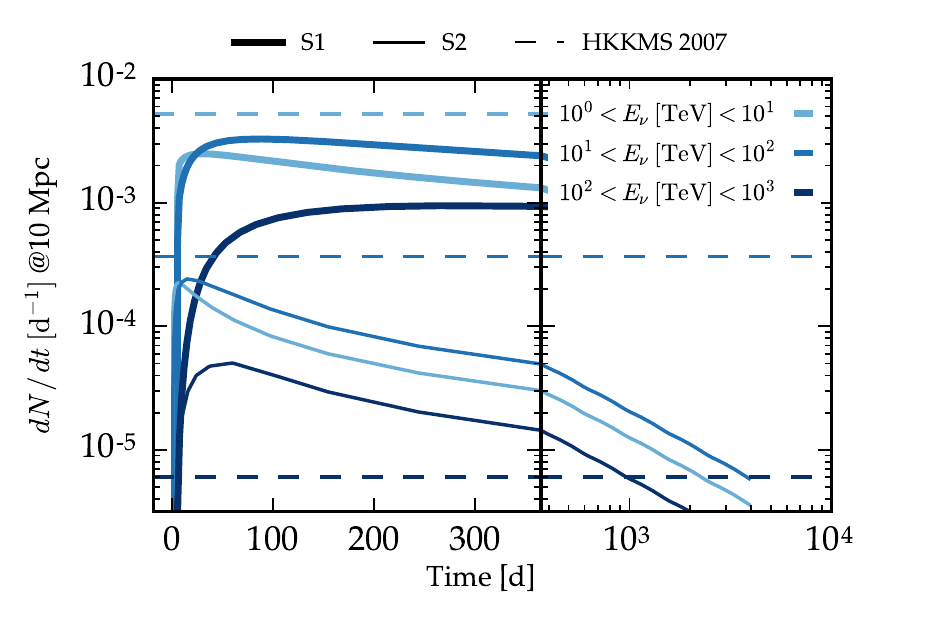}
\caption{Expected IceCube $\nu_\mu+\bar{\nu}_\mu$ event rate as a function of
time for the same parameters as in Fig.~\ref{fig:flux} (S1: thick lines; S2:
thin lines). The curves, in both cases, are truncated at the time the shock
reaches the outer radius of the dense CSM. Different colours correspond to
different energy bands (inset legend).  Dashed lines indicate the atmospheric
neutrino event rate at (from top to bottom) 1-10 TeV, 10-$10^2$ TeV, and
$10^2$-$10^3$ TeV. The event rates are calculated using the IceCube effective
area for up-going muon events \citep[declination -5\textdegree\, to
30\textdegree]{Aartsen:2016oji}).}
\label{fig:rate}
\end{figure}

The expected IceCube $\nu_\mu+\bar{\nu}_\mu$ event rate at different energy
bands and as a function of time is shown in Fig.~\ref{fig:rate} for S1 (thick
lines) and S2 (thin lines).  The rate has been calculated using the effective area of IceCube 
for horizontally up-going muon events \citep[declination
-5\textdegree\, to 30\textdegree]{Aartsen:2016oji}. Dashed lines indicate the
atmospheric neutrino event rate at the same energy bands (see legend). The
event rate for S1 increases rapidly at early times in all energy ranges except
for the $0.1-1$~PeV band. The late-time increase of the rate in this case
($t>50$~d) reflects the increasing maximum proton and neutrino energies with
time. After the peak flux has been reached, the neutrino flux in all energy
bands decays very slowly with time. This indicates that the temporal evolution
of the parent proton distribution is dictated by p-p collisions (see
equations~(\ref{eq:F_pp}) and (\ref{eq:time})). On the contrary, the evolution
of the proton distribution in S2 is governed by adiabatic losses. This is
reflected on the temporal decay of the event rates which scale approximately as
$\propto t^{-1}$ (see  equation~(\ref{eq:F_adiab})). Furthermore, the event
rates in S2 are typically one order of magnitude lower than those otbained in
S1, since ${\rm d}N_\nu/{\rm d}t \propto K_{\rm w}\epr \vsh^3$ and $K_{\rm w}$ in
S2 is ten times lower than in the first scenario (see
Table~\ref{tab:case_parameters}). 

Table~\ref{tab:evt} shows the cumulative event number over a period of 385 days (S1) and 10.7 yr (S2) in different energy bins.
For comparison reasons, the atmospheric neutrino background
for 385~d of exposure in a location of $0.75^\circ$ around the SN location is also shown. 
IceCube's sensitivity on a timescale of $\sim1$~yr (as in S1) translates to a
mean of 3.2 events for an unbroken $E_{\nu}^{-2}$
spectrum~\citep[Fig.~1][]{Aartsen:2015wto} after taking into account the
atmospheric background, while 11 events are required for a 5$\sigma$ discovery.
For $E_{\nu}^{-2}$ spectra that have a sharp cutoff between 100~TeV and 1~PeV,
as in the scenarios discussed here, the sensitivity worsens by a factor of $2$
in flux $E_{\nu}F_\nu$ \citep[Fig.~3][]{Aartsen:2016tpb} or by $36\%$ in total
event count after taking into account the IceCube effective area. Henceforth,
we adopt $N_{\rm s}=4.4$ and $N_{\rm dp}=15$ as the event counts required to
reach IceCube's sensitivity and $5\sigma$~discovery potential, respectively, in
a period of 385~d. We note that these event counts are applicable to the
Northern Sky only, where IceCube selects a pure sample of muon neutrinos. Due
to the cutoff in the neutrino spectrum lying between 100~TeV and 1~PeV, absorption
effects in the Earth can be neglected. 
 \begin{table}
 \centering
 \caption{Expected IceCube $\nu_{\mu}+\bar{\nu}_{\mu}$ events at different
     energy bands from a SN~IIn located at 10 Mpc compared to the number of
     background events. For the S1 and S2 parameters, see text. The source
     events for S1 and S2 are calculated by integrating  the rates shown in
 Fig.~\ref{fig:rate} over a period of 385 days and 10.7 yr, respectively. The
 number of background events is calculated for a period of 385 days over a
 0.75\textdegree \,  circle around the source. } 
 \begin{tabular}{lcccc}
$E_{\nu}$ (TeV)  & \multicolumn{3}{c}{SN~IIn} &Background\\
\hline
              & S1    & \multicolumn{2}{c}{S2} & \\
              &       & {\small First year} & {\small Full duration}     & \\
\hline
0.1-1         & 0.082 & 0.003 & 0.007 & 4.113 \\
1-10          & 0.696 & 0.026 & 0.057 & 2.043 \\
10-$10^2$     & 1.080 & 0.039 & 0.089 & 0.143 \\
$10^2$-$10^3$ & 0.304 & 0.010 & 0.024 & 0.002 \\
0.1-$10^3$    & 2.163 & 0.079 & 0.177 & 6.301 \\
\hline
 \end{tabular}
 \label{tab:evt}
\end{table}

Based on the discussion above and on the results presented in
Table~\ref{tab:evt}), IceCube is able to limit $\epr < 0.06$ for a S1-like
SN~IIn at 10~Mpc distance. The constraint on the proton acceleration efficiency
is more stringent than the one obtained from the analysis of the diffuse
neutrino flux, where $\epr\lesssim 0.2$ for a SNe~IIn rate of $\xi=0.04$ (see
Section \ref{sec:diffuse}).  Were a S1-like SN~IIn detected at $\sim18$~Mpc
from Earth, the point source neutrino analysis would  result in $\epr \lesssim
0.2$, i.e., it would be as constraining as the diffuse neutrino analysis. The
second scenario (S2) is more challenging for neutrino detection due to its
lower neutrino rate. Although the duration in S2 is ten times longer than in
S1, $\sim45\%$ of the signal is expected in the first year
(Table~\ref{tab:evt}). The background of atmospheric neutrinos is, however,
constant in time. Hence, due to the decline in the neutrino rate of S2, the
best signal-to-noise ratio is obtained for a period of $200-400$~d, similar to
the full duration in S1. As a result, IceCube would be able to constrain
$\epr<0.2$, only if a S2-like SN~IIn was detected at $\lesssim
3.4$~Mpc. 

\section{Discussion}
\label{sec:discuss}
SNe surrounded by dense CSM pose an interesting alternative to extragalactic scenarios of high-energy neutrino emission. In this paper, we have calculated the diffuse neutrino emission from this rare class of CC SNe using a Monte Carlo approach that allowed us to incorporate their widely ranging properties, such as mass-loss rates, wind velocities, and proton acceleration efficiencies. We have also demonstrated through two indicative scenarios, the temporal dependence of the neutrino flux from individual SNe and evaluated the possibility of being detected as neutrino point sources with IceCube. In the following, we discuss the main caveats of the model and expand upon several of its aspects, such as possible associations between the IceCube neutrinos and SNe IIn. 

\subsection{Caveats}
The forward shock is expected to be mildly decelerating, unless the SN ejecta have a flat density profile, namely $\rho_{\rm ej} \propto r^{-n_{\rm ej}}$ with $n_{\rm ej}<3$ \citep[e.g.][]{chevalier82}.  For example, $\vsh \propto r^{-0.1}$ for a wind-like CSM ($n\propto r^{-2}$) and ejecta with $n_{\rm ej}=12$. In contrast, the SN shock would propagate with a constant velocity, at least in the free expansion phase, if both media had uniform densities \citep[e.g.][]{matzner99}. A slowly decreasing shock velocity would result in decreasing neutrino flux and maximum neutrino energy, as indicated by equations (\ref{eq:F_pp})-(\ref{eq:F_adiab}) and (\ref{eq:Evmax}), respectively. A faster decrease of the neutrino flux with time could also be caused by the propagation of the SN shock to a CSM with steeper density profile than the one considered here \citep[e.g.][]{chandra15}. Flatter density profiles, on the other hand, would enhance the neutrino production rate. Such deviations have been inferred for a handful 
of SNe IIn detected in radio wavelengths  (e.g. \citet{fransson96, immler01}, but see also \citet{martividal11}).

So far, we have presented results for the high-energy neutrino emission from
SNe IIn shocks propagating in dense CSM. Apart from the forward shock, there is
a second shock wave that forms in the outer parts of the SN ejecta (reverse
shock) and may also contribute to the neutrino signal. Its contribution mainly
depends on its velocity, the proton accelerated energy, and the density of the
shocked SN ejecta. As the maximum energy of accelerated protons and, in turn,
neutrinos scales cubically or quadratically with the shock velocity (see
equation (\ref{eq:Evmax})), small differences between the forward and reverse
shock velocities will lead to large differences in the cutoff energy of the
neutrino spectra (see also Fig.~1 in \citet{murase11}). Furthermore, the
density of the ejecta and the energy of accelerated protons at the reverse
shock affects the overall normalization of the neutrino spectrum. A detailed
calculation of the neutrino emission from both shocks requires hydrodynamical
calculations that are, however, beyond the scope of this paper. Nevertheless,
the contribution of the reverse shock to the total non-thermal (i.e., radio
synchrotron and neutrino)  emission is expected to be small, as long as the
density profile of the ejecta is steep (\citealt{chevalier_fransson03}; see
also Fig.~3 in ZP16). 

While setting up our  Monte Carlo simulations for the diffuse neutrino emission from  SNe IIn, we treated the shock velocity and ejecta mass as independent variables and let the SN kinetic energy ($E_{\rm k}$) to be a derived quantity. This choice may, in principle, lead to extreme values of the SN kinetic energy ($E_{\rm k}\gg 10^{51}$~erg). To check this possible caveat, we computed the distribution of $E_{\rm k}$ values using the generated distributions of the shock velocity and ejecta mass (Section \ref{sec:parameters}). The median SN kinetic energy in our simulations is found to be $5\times 10^{51}$ erg. Moreover, 68 per cent of the simulated sources had kinetic energies between $10^{51}$ and $2\times 10^{52}$~erg, which is not uncommon for SNe IIn; for example, \citet{smith2013} showed that $E_{\rm k} > 10^{51}$ for  2009ip and 2010mc. The small fraction of sources with  $E_{\rm k} \gg 10^{52}$~erg does not affect our main conclusions about the diffuse neutrino emission (Section \ref{sec:diffuse}), especially when other sources of uncertainty are considered, i.e. production rate of SNe IIn (see following paragraph). The injected energy into accelerated protons can also be estimated as $E_{\rm p}=\epsilon_{\rm p}E_{\rm k}$. The distribution of $E_{\rm p}$ derived by our simulations has a median value of $1.5\times 10^{50}$~erg, while 68 per cent of the simulated sources have $3\times10^{49}\le E_{\rm p} \le 8\times10^{50}$~erg. These values should be compared with value $5\times 10^{51}$~erg adopted by ZP16 for all sources.

We have calculated the diffuse neutrino emission from the SN IIn class by
adopting the CC volumetric rate of \citet{hopkins_beacom06} (hereafter, HB06)
and assuming that 4 per cent of all CC SNe are of the type IIn. The rates
predicted by the HB06 model are two times higher than those obtained by the
model of \citet{madau14}--henceforth MD14, at all redshifts
\citep{cappellaro15}. Two recent extragalactic surveys performed with the
Hubble Space Telescope provided volumetric SN rates in high-z galaxies
($z\lesssim 2.5$), which lend further support to the MD14 model
\citep{strolger15}. However, the existing observations do not exclude the HB06
model, especially due to the uncertainties that enter in the model, such as the
mass range of the progenitor stars (see Section 8.1 in \citet{cappellaro15} for
details). The rate of SNe with dense and massive CSM is also uncertain. Here,
we adopted $\xi=0.04$, but this could range from 1 per cent to $\sim$6 per cent
\citep[e.g.][]{smith14, strolger15, cappellaro15}.  Adoption of the MD14 model
with $\xi\sim 0.08$ would, thus, lead to similar results for the diffuse
neutrino emission as those presented in Fig.~\ref{fig:diffuse}. 

The largest uncertainties ($\sim50$ per cent) entering in the estimation of the
diffuse neutrino flux are those described above. If the SNe~IIn rate was lower by a
factor of two, IceCube HESE observations would limit the proton accelerated energy fraction at 
$\lesssim0.45$, or $\epr\lesssim0.3$ after including the IceCube up-going muon neutrino flux above 195~TeV.
\begin{figure}
\centering
\includegraphics[width=0.49\textwidth,trim=30 10 0 0]{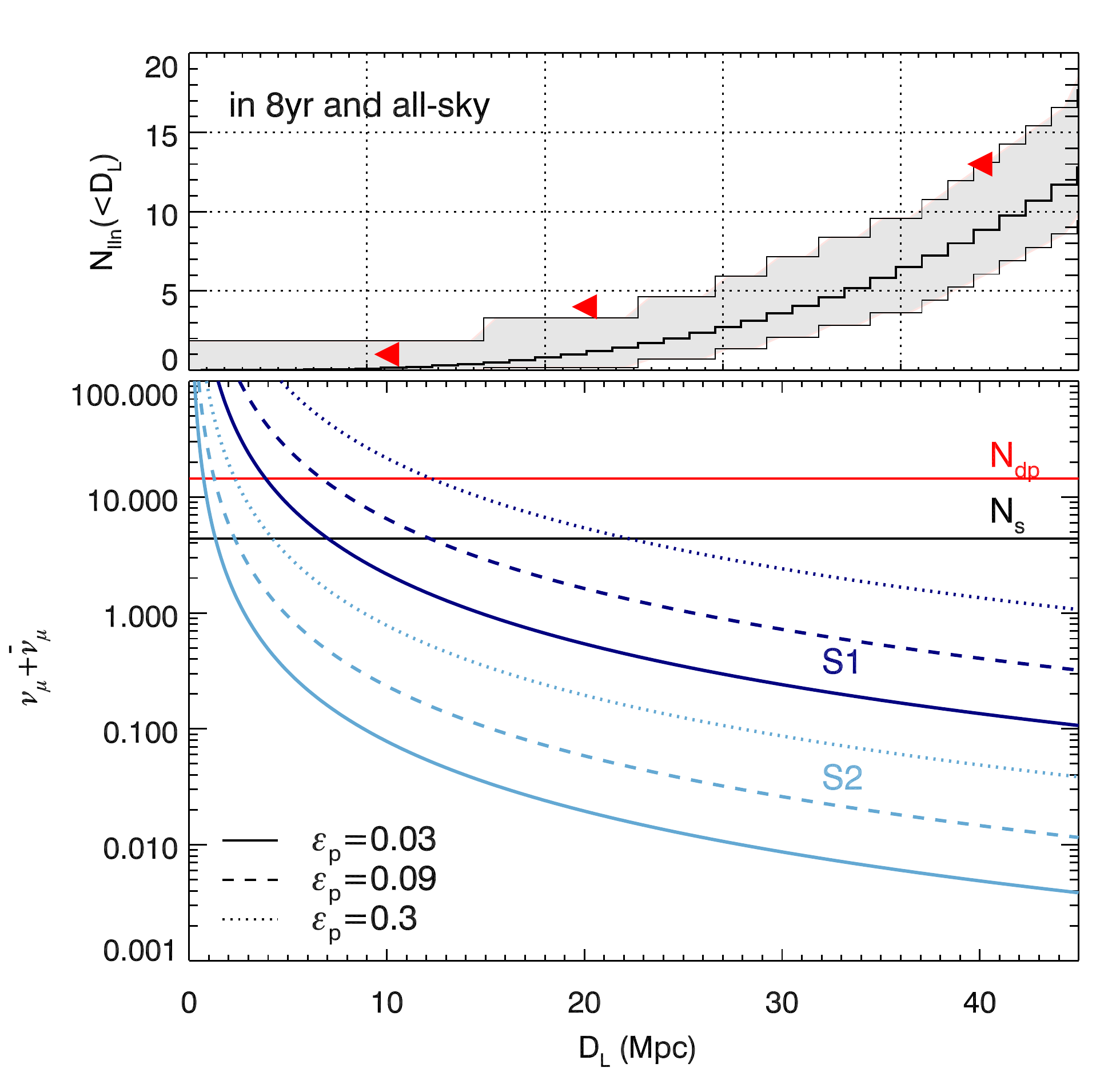}
\caption{Bottom panel: Expected IceCube $\nu_\mu+\bar{\nu}_\mu$ event number as
a function of luminosity distance $D_{\rm L}$ for the two SN scenarios
discussed in text (S1: blue lines; S2: cyan lines).  For the second scenario,
the number of neutrinos expected only within the first year is shown. For both
scenarios, the results for three different values of the proton acceleration
efficiency are presented (see legend). The horizontal black and red lines
indicate the number of  neutrinos required to reach IceCube's sensitivity and
$5\sigma$~discovery potential, respectively. Top panel: All-sky number of
SNe~IIn up to $D_{\rm L}$ expected over 8 years (black thick line). The shaded
region corresponds to the 84 per cent confidence level (CL), assuming that the
supernova number follows a Poissonian distribution \citep{gehrels86}. Red
coloured symbols denote the observed number of SNe~IIn within that distance
(see Table~\ref{tab:tab2}).}
\label{fig:dist_ep}
\end{figure} 
\subsection{Constraints on $\epr$}
Numerical simulations of particle acceleration at non-relativistic shocks are
the main means of deriving the proton acceleration efficiency $\epr$ from first
principles \citep{caprioli_spitkovsky14b, park2015}. Additional constraints on
the acceleration efficiency can be placed through the detection (or
non-detection) of  multi-messenger signatures associated with relativistic
protons. In this study, we have demonstrated how high-energy neutrino
observations may be used to assess $\epr$ at SN shocks propagating in dense
CSM.  In particular, the IceCube's measurement of the diffuse neutrino flux 
already constrains $\epr \lesssim 0.3\times(0.04/\xi)$ or $\epr\lesssim0.2 \times(0.04/\xi)$ when including the up-going muon neutrino sample from
the Northern Sky (Section \ref{sec:diffuse}). Even more stringent constraints on the
acceleration efficiency may be placed by direct observations of close-by
SNe~IIn with neutrino telescopes (Section \ref{sec:pointsource}). 

The neutrino flux from a SN~IIn depends on its distance, shock velocity,
mass-loading parameter, and proton acceleration efficiency (Section
\ref{sec:pointsource}). All parameters, except for the latter, can be, in
principle, inferred from radio and optical observations of SNe~IIn.  For
example, $\vsh$ is routinely inferred from the width of optical emission lines,
while $K_{\rm w}$ can be derived by the fitting of radio observations
\citep{chevalier98,chevalier06,chandra15}. Besides the intrinsic source
parameters that affect the neutrino luminosity, as exemplified through
scenarios S1 and S2, the actual distance of the source has the strongest impact
on the IceCube expected neutrino rate (see equation~(\ref{eq:Fv})). 

\begin{table*}
 \centering 
 \caption{List of SNe~IIn detected within 40 Mpc over a period of eight years (2008-2015).}
\begin{tabular}{ccc ccc ccc}
\hline
Name    &   $T_{\max}$ (MJD)\textdagger &   Host Galaxy    &      R.A.  (J2000)    &  Dec.  (J2000)     &    z      & $D_{\rm L}$ (Mpc)   & Type   & 	References\\
\hline
SN2008S    & 54508.5 &	NGC 6946     &  20:34:45.4   &  +60:05:57.8  & 0.0002 &       0.7   &IIn-pec/LBV    &  (1)     \\  
SN2008X    & 54502.5 &	NGC 4141     &  12:09:48.33  &  +58:51:01.6  & 0.006  &     28.0    &IIn & (2)  \\  
SN2009ip\textdaggerdbl   & 56206.5 &	NGC 7259     &   22:23:08.3  &   -28:56:52.4 &    0.006 &     26.4 &IIn  & (3)      \\
SN2009kr   & 55148.5 &	NGC 1832     &   05:12:03.3  &   -15:41:52.2 &    0.007   &      16.0    &II             &  (4)       \\ 
SN2011A    & 55562.5 &	NGC 4902     &   13:01:01.2  &   -14:31:34.8 &   0.009 &     39.7 &IIn            &   (5)     \\
SN2011fh   & 55797.5 &	NGC 4806     &   12:56:14.0  &   -29:29:54.8 &   0.008    &       36.0     &IIn            & (6)    \\ 
SN2011ht   & 55879.5 &	UGC 5460     &   10:08:10.58 &   +51:50:57.1 &   0.004   &       16.0     &IIn            &  (7)     \\  
SN2013gc   & 56603.5 &	ESO 430-G20  &   08:07:11.9  &   -28:03:26.3 &   0.003 &       15.1 &IIn            &  (8)   \\ 
PSN J14041297-0938168& 56645.5 &IC 4363& 14:04:13.0   & -09:38:16.8 &  0.003 &       12.5  &IIn            & (9)     \\ 
CSS140111:060437-123740& 56734.5	& MCG-02-16-02 & 06:04:36.71 & -12:37:40.6 & 0.007 &  32.9  &IIn  & (10)      \\  
SN2014G    &  56674.5 &	NGC 3448     &   10:54:34.13 &   +54:17:56.9 &   0.005 &   20.0  &II L           & (11)    \\ 
SN2015bh   &  57163.5 &	NGC 2770     &   09:09:34.96 &   +33:07:20.4 &   0.006  &      28.5   &Impostor       &   (12)   \\  
SN2015J    &  57201.5 &	A073505-6907 &   07:35:05.2  &   -69:07:53.1 &    0.005  &      24.0     &IIn            & (13)      \\
PSN J13522411+3941286& 57071.5	&	NGC5337 & 13:52:24.1  & +39:41:28.6 & 0.007 &      32.1  &IIn            & (14)   \\  
ASASSN-15lf& 57194.5	&	NGC 4108   & 12:06:45.56      & +67:09:24.00 &  0.008 &      37.3  &IIn            &  (15) \\ 
SNhunt248  & 56828.5 &	NGC 5806     &   14:59:59.5  &   +01:54:26.2 &    0.005 &     20.1 &IIn Pec        &   (16)    \\       
\hline  
\end{tabular}
\\ \textdagger Time of maximum optical light. \textdaggerdbl Available radio light curve. No radio observations exist for the rest of the sources. 
(1): \cite{Arbour2008,Botticella2009,Brown2014}; (2):\cite{Chandra2008}; (3): \cite{Maza2009,Fraser2013,Potashov2013,Mauerhan2013,Pastorello2013,Graham2014,Margutti2014,Brown2014,Fraser2015}; (4): \cite{Nakano2009}; (5): \cite{deJaeger2015,Stritzinger2011}; (6):  \cite{Prieto2011}; (7):  \cite{Pastorello2011,Humphreys2012,Mauerhan2013b,Brown2014,Ofek2014}; (8): \cite{Antezana2013};
(9): \url{http://www.cbat.eps.harvard.edu/unconf/followups/J14041297-0938168.html}; (10): \url{http://nesssi.cacr.caltech.edu/catalina/current.html}; (11): \cite{Terreran2016,Denisenko2014}; (12): \cite{Elias-Rosa2015, Elias-Rosa2016}; (13): \cite{Childress2015};  (14): \cite{Zhang2015}; (15): \cite{Masi2015,Challis2015}; (16): \cite{Mauerhan2015,Mauerhan2017,Kankare2015} 
 \label{tab:tab2}
\end{table*}  
The IceCube $\nu_\mu+\bar{\nu}_\mu$ event number expected  within the first
year in both scenarios discussed in Section \ref{sec:pointsource}, is plotted
in Fig.~\ref{fig:dist_ep} (bottom panel) as a function of the source distance.
Different types of lines correspond to three values of the proton acceleration
efficiency marked on the plot. The curves are obtained after scaling the total
number of muon neutrinos obtained in S1 and S2 (see Table~\ref{tab:evt}) with
the source distance and $\epr$. The top panel in Fig.~\ref{fig:dist_ep} shows
the all-sky number of SNe~IIn expected over 8 years up to a distance $D_{\rm
L}$ (black thick line); the number is derived using the CC SN rate of HB06 and $\xi=0.04$. 
For $\epr$ values allowed by the diffuse flux measurements, i.e., $\epr
\lesssim 0.2 \times (0.04/\xi)$, only S1-like SNe~IIn at distances
$\lesssim18$~Mpc are strong enough neutrino emitters to constrain less
efficient acceleration scenarios, in agreement with the predictions by \cite{murase11}. If $\epr=0.2$,
IceCube would be capable of claiming a neutrino detection from an S1-like
SN~IIn exploding at $\sim 10$ Mpc. For SNe~IIn with lower mass-loading parameter $K_{\rm w}$, as discussed in S2,
the neutrino rate decreases significantly, even for $\epr=0.2$. This restricts
the accessible distance to no more than 4~Mpc. 

S1-like sources with faster
shocks are more promising candidates for neutrino detection. For higher shock
velocities, the neutrino rate increases as $\propto \vsh^3$ (see equation
(\ref{eq:F_pp})), while the duration of the neutrino production decreases with
$\vsh^{-1}$. The latter leads to an accordingly lower number of background
events, thus increasing the signal-to-noise. Furthermore, the higher shock
velocity results in an increased cutoff value of the neutrino energy spectrum,
as this scales with $\vsh^2$ (Section \ref{sec:pointsource}). For example, a
source with a three times higher shock velocity than in S1
($\vsh=3\times10^4$~km~s$^{-1}$) and the same $K_{\rm w}, \epr$ values,  would
yield $\sim 38$ events above $10$~TeV. Hence, IceCube would be sensitive to
such a source up to 40~Mpc even for $\epr=0.03$. A positive detection would
also be possible for a source located at $\lesssim 20$~Mpc. Given that
IceCube's sensitivity improves when the neutrino spectrum extends above 100 TeV
\citep[Fig. 3 in][]{Aartsen:2016tpb}, these estimates are rather conservative.  

Another important aspect in the search of neutrino point sources is the rate of
SNe~IIn explosions at the relevant distances (i.e., $\lesssim 40$~Mpc).  The
expected number of SNe~IIn at distances $\lesssim 22$~Mpc (40 Mpc) is
$0.7^{+0.9}_{-0.5}$ ($4.4^{+2.1}_{-1.4}$)\footnote{The errors correspond to 1
$\sigma$ errors of Gaussian statistics.} within 8 years in the Northern Sky
(top panel in Fig.~\ref{fig:dist_ep}).  During the period of 2010-2016 three
SNe~IIn were detected in the Northern Sky at a mean distance of $\sim29$~Mpc
(Table~\ref{tab:tab2}). Stacking of their neutrino signal could place stronger
constraints on $\epr$ than the diffuse neutrino flux by a factor of
$\sqrt{N_{\rm IIn}}$.

Neutrino production in p-p collisions is accompanied by the injection of
relativistic electron-positron pairs in the post-shock region and the
production of GeV $\gamma$-ray photons via the decay of neutral pions. The
first systematic search for $\gamma$-ray emission in \fermi  \, data from the
ensemble of 147 SNe~IIn exploding in a dense CSM was recently presented by
\citet{ackermann15}. No significant excess above the background was found
leading to model-independent $\gamma$-ray flux upper limits. These $\gamma$-ray
non-detections constrained the ratio of the $\gamma$-ray to optical luminosity
in the range 0.01-1. These can, in principle,  be translated to limits on
$\epr$. However, due to the uncertainty in the escaping fraction of
$\gamma$-rays, no stringent limits could be placed on the proton acceleration
efficiency. 

\subsection{IceCube neutrinos in coincidence with SNe~IIn?}
Table~\ref{tab:tab2} shows the detected SNe~IIn in the local Universe (i.e.,
within a distance $D_{\rm L}=40$~Mpc). Out of the 29 high-energy muon neutrinos
detected in the Northern Sky above 200~TeV \citep{Aartsen:2016xlq}, none is
found to be in spatial or temporal coincidence with these SNe~IIn. However, one
cascade-like event\footnote{ID~16 with deposited energy of 30.6~TeV and
19.4~degrees median angular uncertainty.} from the 4-year HESE IceCube sample
\citep{ICRC15} was detected on MJD~55798.63, i.e. $\delta t=1.13$~days after
the peak of the SN2011fh optical light curve, 
with an angular offset of 7.12~degrees.\footnote{This neutrino event has
another possible connection to SN2011A which was detected 236.13~days prior to
the neutrino at an angular separation of 8.61~degrees.} 

The coincident rate of SNe~IIn in the local Universe ($z\ll 0.1$) during the
time window $\delta t$ and within the positional uncertainty $\sigma_\nu$ of a
neutrino event can be estimated as:
\eqb
    N_{\rm IIn}&=&\int\limits_0^{ D_{\rm L}} {\rm d}r\,r^2
        \sum_\nu\int\limits_{t_\nu-\delta t}^{t_\nu}{\rm d}t\,
        \int\limits_{\Delta\Omega_\nu}{\rm d}\Omega\,\dot{n}_{\rm IIn}\\ \nonumber 
    &=&\xi \, \dot{n}_{\rm SNII}\times\frac{D_{\rm L}^3}{3}\times\delta t \times\sum_\nu 2\pi\left(1-\cos\sigma_\nu\right).
    \label{eq:SN_number}
\eqe
For a IIn fraction of $\xi=0.04$ and a local rate of CC SNe
$10^{-4}$~Mpc$^{-3}$~yr$^{-1}$, the observation of the cascade event ID~16
deviates from the background expectation (i.e., $N_{\rm IIn}=0$ in $\delta t$)
at $2.76 \, \sigma$.  A similar result was obtained after scrambling the
neutrinos of the 4-year HESE sample in right ascension and time or the SNe~IIn
(Table~\ref{tab:tab2}, in direction and time if detected within IceCube
livetime) and counting the number of random associations of one SN~IIn at an
angular distance smaller than the neutrino uncertainty and at a time $\delta
t<1.13$~days. If the cascade-like event was indeed physically associated to SN2011fh,
additional muon neutrino events should have been detected by the ANTARES
neutrino telescope, as this has three to four times larger effective area at
30~TeV than IceCube for up-going muon neutrino events
\citep{Adrian-Martinez:2014wzf}.

Not much information is available for SN2011fh besides the velocity of its
ejecta $v_{\rm ej} =7400$ km s$^{-1}$, which is a good indicator of the shock
velocity  ($\vsh \gtrsim v_{\rm ej}$).  Inspection of
Fig.~\ref{fig:dist_ep} shows that neither IceCube nor ANTARES would be
sensitive to neutrinos from SN2011fh located at 36 Mpc, if this had the same
properties as the source in scenario S1. In addition, ID~16 is an event
produced within the detector (member of the HESE sample). The rate of such
events is much lower (by a factor $\sim 180$ at 30 TeV) than the rate of
up-going muon neutrinos (Table~\ref{tab:evt}), which was used in producing
Fig.~\ref{fig:dist_ep}.  Regardless, a physical association of SN2011fh with
the cascade event ID~16 would require higher $K_{\rm w}$ or/and $\epr$ than
those used in S1. 
 
 An estimate of the mass-loading parameter and proton acceleration efficiency
 in SN2011fh can be derived assuming a physical connection with the cascade
 event ID~16 and using our estimates of the expected neutrino rate. In the
 energy range of 10--100~TeV, we expect $N_{\rm S1}\sim1$ up-going muon
 neutrino events (see Table~\ref{tab:evt}). One neutrino (ID 16) was observed
 in the starting event channel, which has a $\sim200$ times smaller signal
 expectation in the energy range 15--45~TeV. The $90\%$ limits for the
 detection of one starting event with no background is  $N_{\rm
 2011fh}=0.11\dots4.36$ \citep{Feldman:1997qc}. Assuming a year-long duration,
 the product $K_{\rm w}\times \epr$ for SN2011fh can be estimated as (see also
 equation~(\ref{eq:F_pp})):
\eqb
\frac{\left.K_w\times\epr\right|_{\rm 2011fh}}{\left.K_w\times\epr\right|_{\rm S1}}
& = & 200 \left(\frac{v_{\rm sh}^{\rm S1}}{v_{\rm sh}^{\rm 2011fh}}\right)^{3}
\left(\frac{D_L^{\rm 2011fh}}{D_L^{\rm S1}}\right)^{2}\frac{N_{\rm 2011fh}}{N_{\rm S1}} \\  \nonumber
& \approx& 579\dots2.3 \times 10^4.
\eqe
Assuming that in SN2011fh $\epr=0.2$ (i.e., the largest allowed value from the
diffuse neutrino flux measurements), its mass-loading parameter has to be at
least $K_{\rm w}^{\rm 2011fh}\sim 87\, K_w^{\rm S1}\sim 4\times 10^{18}$~g
cm$^{-1}$.  Radio observations of SN2011fh could place strong constraints on
$K_{\rm w}^{\rm 2011fh}$ and, ultimately, exclude (or not) a physical
connection to neutrino ID~16.

\subsection{Diffuse $\gamma$-ray emission}
The neutrino emission resulting from the SN~IIn class is accompanied by a
diffuse $\gamma$-ray component, which may contribute to the isotropic
$\gamma$-ray background (IGRB) at GeV energies due to cascades initiated by the
absorption of multi-TeV photons on the extragalactic background light
\citep[e.g.][]{murase12}. Assuming that the sources are optically thin to
photon-photon absorption, \citet{murase13} showed that generic p-p scenarios of
neutrino production with spectra ${\rm d}N_{\nu}/{\rm d}E_{\nu}\propto
E_{\nu}^{-2}$, for $E_{\nu} \le E_{\nu, \rm b} \sim 1$~PeV and $\propto
E_{\nu}^{-p}$ otherwise, could explain IceCube observations above 100 TeV, if
$p\lesssim 2.1-2.2$.  In general, tighter constraints on the power-law index of
the relativistic proton distribution can be obtained (i.e., $p \sim 2$), if
$E_{\nu, \rm b}\lesssim 30$~TeV. The induced $\gamma$-ray emission from generic
p-p scenarios of neutrino emission was shown to be marginally consistent with
the IGRB \citep[][]{bechtol15, murase16}, if $p=2.5$ and the break energy was
fixed to the lowest energy bin of the combined neutrino data between 25 TeV and
2.8 PeV \citep{aartsen15_combined}. The tension with the IGRB can be relaxed if
a fraction of the multi-TeV $\gamma$-rays is absorbed in the source
\citep[][]{murase16}, which is not unexpected for the early-time evolution of
SNe with dense CSM \citep{kantzas16}. An alternative way of reconciling the p-p
scenario with the IGRB measurements would be $E_{\nu, \rm b}>25$~TeV, which is
still consistent with the neutrino data. 

We showed that the diffuse neutrino flux (per flavour) from SNe~IIn is $\sim
4\times 10^{-9}$~GeV cm$^{-2}$ s$^{-1}$ sr$^{-1}$ or  $\sim 10$ per cent of the
observed IceCube flux above 100 TeV, if the proton acceleration efficiency
varies between 0.01 and 0.1 among the sources (Section \ref{sec:diffuse}).
Assuming no attenuation of multi-TeV $\gamma$-rays in the sources, the diffuse
$\gamma$-ray flux from SNe~IIn at $\sim 50$ GeV is expected to be $\sim
10^{-8}$~GeV cm$^{-2}$ s$^{-1}$ sr$^{-1}$ \citep[see also Fig.~4
in][]{bechtol15}. Only if all shocks in SNe~IIn had the same acceleration
efficiency $\epr \sim 0.2 \times (0.04/\xi)$, could the SN~IIn class explain the
$>100$~TeV neutrino observations. In this case, the accompanying diffuse
$\gamma$-ray emission would be marginally consistent with the IGRB, but it
would exceed the non-blazar contribution to the EGB \citep{bechtol15,
ackermann16}.

\section{Conclusions}
\label{sec:conclusions}
We have evaluated the possible neutrino signal of SNe~IIn and placed
constraints on the proton accelerated energy fraction, $\epr$, by means of
diffuse and point-source neutrino observations with IceCube. By employing a
Monte Carlo method that takes into account the wide spread in the physical
properties of SNe~IIn, we showed that the diffuse neutrino emission from SNe
IIn can account for $\sim 10$ per cent of the observed IceCube neutrino flux
above 100 TeV. In the less realistic scenario, where the proton accelerated
energy fraction is the same for all SN shocks, we showed that the observed
diffuse astrophysical neutrino spectrum could be explained totally by SNe~IIn,
if $\epr \lesssim 0.2$ and 4 per cent of the CC SNe were of the IIn type.
However, the identification of a single  SN~IIn as a neutrino point source with
IceCube using up-going muon neutrinos could place stronger constraints on
$\epr$.  We concluded that such an identification is possible in the first year
following the SN shock breakout for sources within $\lesssim$18~Mpc and $\epr
\lesssim 0.2$ or $\lesssim 7$~Mpc and $\epr \lesssim 0.03$.
Interestingly, one cascade-like event (ID~16) from the 4-year HESE sample of
IceCube was found to be in spatial agreement with SN2011fh ($D_{\rm L}=$36 Mpc)
and was detected only 27.12~hours after the maximum optical SN light. The
probability that this association was not a chance coincidence was found to be
$2.76 \, \sigma$. In case of a positive connection, additional muon neutrinos
should be detected by ANTARES in coincidence with SN2011fh, which should have a
very high mass-loading parameter ($>3.9\times 10^{18}$ g cm$^{-1}$).
Analysis of propitiatory ANTARES data and of SN2011fh radio observations are
strongly encouraged to resolve the nature of this association.

\section*{Acknowledgments} 
We thank the anonymous referee, E. Resconi, and K. Kotera for insightful comments on the manuscript. We also thank K. Murase and A. Mastichiadis 
for useful discussions.  We thank A. Hopkins for providing the theoretical curves of the volumetric
production rate of Type II supernovae. SC is supported by the cluster of
excellence ``Origin and Structure of the Universe'' of the Deutsche
Forschungsgemeinschaft.  GV acknowledges support from the BMWi/DLR grant FKZ 50
OR 1208.

\bibliographystyle{mnras}
\bibliography{sn_neutrinos.bib} 
\end{document}